\documentclass{jfm}
\usepackage{graphicx}
\usepackage{epstopdf, epsfig}
\usepackage{amsmath}
\usepackage{xfrac}
\usepackage[dvipsnames]{xcolor}
\usepackage{textcomp}
\usepackage[normalem]{ulem}
\usepackage{amssymb}

\newcommand{\beginsupplement}{%
        \setcounter{table}{0}
        \renewcommand{\thetable}{S\arabic{table}}%
        \setcounter{equation}{0}
        \renewcommand{\theequation}{S\arabic{equation}}
        \setcounter{figure}{0}
        \renewcommand{\thefigure}{S\arabic{figure}}%
        }

\graphicspath{{./final_figures_GOOD/}}        
        
\usepackage{xcolor}

\shorttitle{Convective dissolution of CO$_2$ in 2D and 3D: impact of hydrodynamic dispersion}

\shortauthor{J. Dhar, P. Meunier, F. Nadal and Y. M{\'e}heust}

\title{Convective dissolution of CO$_2$ in 2D and 3D porous media: the impact of hydrodynamic dispersion}

\author{Jayabrata Dhar\aff{1}
 Patrice Meunier\aff{2}
 Fran\c{c}ois Nadal\aff{3}
 \and Yves M{\'e}heust\aff{1}
 \corresp{\email{yves.meheust@univ-rennes1.fr}}
}

\affiliation{\aff{1}Univ. Rennes, CNRS, G{\'e}osciences Rennes (UMR6118), 35042 Rennes, France
\aff{2}Aix Marseille Universit{\'e}, Centrale Marseille, CNRS, IRPHE, 13384 Marseille, France
\aff{3}Wolfson school of Mechanical and Electrical Engineering, Loughborough University, Loughborough, UK
}

\begin{document}

\maketitle

\begin{abstract}
Subsurface storage of CO$_2$  is widely regarded as the most promising measure to limit global warming of the Earth's atmosphere. Convective dissolution is the process by which CO$_2$ injected in deep geological formations dissolves into the aqueous phase, which allows storing it perennially by gravity. The process can be modeled by buoyancy-coupled Darcy flow and solute transport. The transport equation should include a diffusive term accounting for hydrodynamics (or, mechanical) dispersion, with an effective diffusion coefficient that is proportional to the local interstitial velocity. A few two-dimensional (2D) numerical studies, and three-dimensional (3D) experimental investigations, have investigated the impact of hydrodynamic dispersion on convection dynamics, with contradictory conclusions drawn from the two approaches. Here, we investigate systematically the impact of the strength $S$ of hydrodynamic dispersion (relative to molecular diffusion), and of the anisotropy $\alpha$ of its tensor, on convective dissolution in 2D and 3D geometries. We use a new numerical model and analyze the following quantities:  the solute fingers' number density (FND), penetration depth and maximum velocity; the onset time of convection; the dissolution flux in the quasi-constant flux regime; the mean concentration of the dissolved CO$_2$; and the scalar dissipation rate. We observe that for a given Rayleigh number ($Ra$) the efficiency of convective dissolution over long times is mostly controlled by the onset time of convection. For porous media with $\alpha = 0.1$, commonly found in the subsurface, the onset time is found to increase as a function of dispersion strength, in agreement with previous experimental findings and in stark contrast to previous numerical findings. We show that the latter studies did not maintain a constant $Ra$ when varying the dispersion strength in their non-dimensional model, which explains the discrepancy. When considering larger $\alpha$ values, the dependence of the onset time on $S$ becomes more complex: non-monotonic at intermediate values of $\alpha$, and monotonically decreasing at large values. Hence hydrodynamic dispersion either slows down convective dissolution (in most cases) or accelerates it depending on the anisotropy of the dispersion tensor. Furthermore, systematic comparison between 2D and 3D results show that they are fully consistent with each other on all accounts, except that in 3D the onset time is slightly smaller, the dissolution flux in the quasi-constant flux regime is slightly larger, and the dependence of the FND on the dispersion parameters is impacted by $Ra$.
\end{abstract}




\section{Introduction}

It is now widely admitted that the global warming of the Earth's atmosphere observed since the beginning of the industrial era, in particular in the last 30 years, mostly results from an increase in the concentration of atmospheric greenhouse gases, among which CO$_2$ accounts for about two thirds of the temperature increase \citep{IPPCrep2013}. In this context, the storage of CO$_2$ in deep geological formations (deep saline aquifers and depleted oil/gas reservoirs)  is widely regarded as the most promising measure to limit global warming, and has thus attracted much attention from the scientific and engineering communities. Upon injection at depths larger than $900$ m, CO$_2$ is in its supercritical state (scCO$_2$), 
where it is less dense than the resident brine, so it rises towards the top of the geological formation where it is trapped by an impermeable cap rock. At the scCO$_2$-brine interface, scCO$_2$ partially dissolves into the brine, thereby forming an aqueous layer of brine enriched with dissolved CO$_2$, that is densier than the brine below. Hence this unstable stratification of the two miscible fluids (scCO$_2$ and brine) leads to a gravitational instability wherein the ensuing convection allows the dissolved sCO$_2$ to be transported deeper into the formation, while fresh CO$_2$-devoid brine is brought to the scCO$_2$-brine interface, allowing CO$_2$ to further dissolve into the aqueous phase \citep{daniel2013optimal,huppertAnnuRevFluidMech2014,tilton2014nonlinear}. This so-called solubility trapping mechanism allows for perennial trapping of CO$_2$ by gravity within the resident brine.  The total storage capacity is constrained by the available porous volume and solubility of scCO$_2$ into the brine. 

But the flow dynamics at play until that capacity is reached is complex. It initiates with the formation of a transient diffusive layer of brine-CO$_2$ mixture that strongly damps small perturbations until a critical time is reached. After that time, the gravitational instability develops, first in the linear regime where the growth of perturbations to the miscible interface is exponential \citep{riaz2006onset,tilton2013initial}. In this regime  the total diffusive flux at the top boundary of the flow domain (also equal to the total dissolution flux as no convective flux exists on that boundary) decreases while the convective perturbations grow. The strength of the convection (with respect to molecular diffusion) is quantified by the non-dimensional Rayleigh number. The time at which the total dissolution flux starts increasing again, denoted the nonlinear onset time, is that at which the convective process can be considered to develop. The total flux then reaches a (quasi-)constant flux regime in which it fluctuates around a plateau value which is all the better defined as the Rayleight number is larger \citep{emami2015convective}. The so-called shutdown regime that follows results from the progressive rise in the mean solute CO$_2$ concentration in the brine, which weakens the convection.

Understanding this dynamics is crucial for the prediction of the characteristic time scale of solubility trapping in particular, and of the subsurface sequestration of CO$_2$ in general. Hence convective dissolution has been the topic of many past studies, either based on experiments in Hele-Shaw cells (i.e., setups analog to a homogeneous two-dimensional -- 2D -- porous medium) \citep{vremePRF2016} or in granular porous media imaged using optical methods  \citep{liang2018effect,brouzet2021SUBMITTED} or X-ray tomography \citep{wang2016three,liyanage2019multidimensional} \cite{Cheng2012WRR}, or on theoretical/numerical simulations \citep{hidalgo2009effect,neufeld2010convective,elenius2012time,tilton2013initial}, to name a few. Among the latter, only a handful \citep{Pau2010threeD,Hewitt2014HighRa,Fu2013Pattern3D,Green2018Heterogeneous3D} have presented three-dimensional (3D) simulations.

Not considering potential chemical reactions between the dissolved CO$_2$ and either the solid phase or other solutes, Darcy-scale theoretical modeling of convective dissolution describes the buoyancy-driven hydrodynamics and transport of dissolved CO$_2$, coupled through the dependence of the brine's density on the local concentration in dissolved CO$_2$. At the Darcy scale, solute diffusion does not only result from molecular diffusion, but also from hydrodynamic (or, mechanical) dispersion, which is the Darcy-scale manifestation of the pore scale interaction between molecular diffusion and heterogeneous advection \citep{FRIED1971}. A well-posed formulation of the transport equation should thus include a diffusive term accounting for hydrodynamic dispersion, i.e., involving a dispersion tensor that is proportional to the magnitude of the local velocity vector as well as anisotropic, since mechanical dispersion is typically one order of magnitude larger along the local velocity's direction than along the transverse direction \citep{perkins1963review,bijeljic2007pore}; the dispersivity length which define the linear transformation to be applied to the local velocity are intrinsic properties of the porous medium.  Such a dispersive term can potentially turn the dynamics highly non-linear. 

In effect,  many of the aforementioned numerical studies have considered simple diffusive transport, either because considering only molecular diffusion allows for analytical developments otherwise intractable, or because they considered a Péclet number sufficiently small  for molecular diffusion to always be negligible (this hypothesis being in any case reasonable at the initiation of the instability). In particular, to our knowledge, those among the 3D numerical studies which have accounted for hydrodynamic dispersion have not investigated its effect in detail, but have rather focused on the spreading of buoyant current \citep{Marco2021current} and the effect of carbonate geochemical reactions \citep{Erfani2021Reaction3D}. Furthermore, while the nonlinear onset time is known from numerical and experimental studies alike to decrease with the Rayleigh number  \citep{riaz2006onset,daniel2013optimal,tilton2014nonlinear,liyanage2019multidimensional},  the literature has reported conflicting views regarding he influence of hydrodynamic dispersion on the onset of gravitational instability  \citep{wen2018rayleigh,Marco2021current}. Numerical studies \citep{hidalgo2009effect,ghesmat2011effect} have predicted that the nonlinear onset time decreases monotonically with the strength of hydrodynamic dispersion, and that this reduction may reach two orders of magnitude. On the contrary, the theoretical/numerical by \citet{emami2017dispersion} reported an increase of the onset time when the dispersion strength is increased, and experimental studies \citep{menand2005dispersion,liang2018effect} have reported an evolution of the convection's structure when increasing the dispersion strength that also indicates a weakening of convection. Moreover, the impact of the dispersion tensor's anisotropy has no been studied systematically, though the numerical predictions of \citep{ghesmat2011effect,xie2011speed} indicate that its overall effect on the efficiency of convective dissolution is negligible.

In this study, we use a new in-house numerical model accounting for anisotropic hydrodynamic dispersion and implemented using the open-source  OpenFOAM computational fluid dynamics (CFD) platform OpenFOAM to investigate convective dissolution in 2D and 3D geometries. Our objectives are two-fold. Firstly, to investigate systematically how the strength of hydrodynamic dispersion (relative to molecular diffusion), and the anisotropy of its tensor impact convective dissolution. In doing so we explain the discrepancy between the conclusions drawn on the role of hydrodynamic dispersion by previous numerical and experimental studies; our own numerical results are consistent with the experimental observations. Secondly, to compare systematically the results obtained in the 2D and 3D geometries, all parameters being equal otherwise, to assess the role of space dimensionality on model predictions. The numerical model is based on a stream function formulation. The parameter space is investigated as widely as possible given the large computational times associated in particular to 3D  numerical simulation: three dispersion strengths, three to four dispersivity length anisotropies, two Rayleigh numbers. The assessment of the impact of hydrodynamic dispersion is based on all available physical observables:  the solute fingers' number density (FND), penetration depth and maximum velocity; the onset time of convection; the dissolution flux in the quasi-constant flux regime; the mean concentration of the dissolved CO$_2$; and the scalar dissipation rate.

The theoretical model and its numerical implementation are described in section~\ref{sec:model}. The systematic investigation of the dependence of all aforementioned observables on the dispersion strength and dispersivity length anisotropy is presented in section~\ref{sec:results}. In the Discussion (section~\ref{sec:discussion}) we provide a synthetis of our findings on the role of dispersion and on that of space dimensionality, and confront these results to those of previous studies on the topic; we also discuss the role of space dimensionality. Finally the Conclusion presents a short summary of the content of the paper,  a synthesis of its main finding, and some  prospects for future studies.

\section{Model of convective dissolution in a homogeneous porous medium}
\label{sec:model}

\subsection{Theoretical Formulation and boundary conditions}

\begin{figure}
\centering
\includegraphics[width=0.75\textwidth]{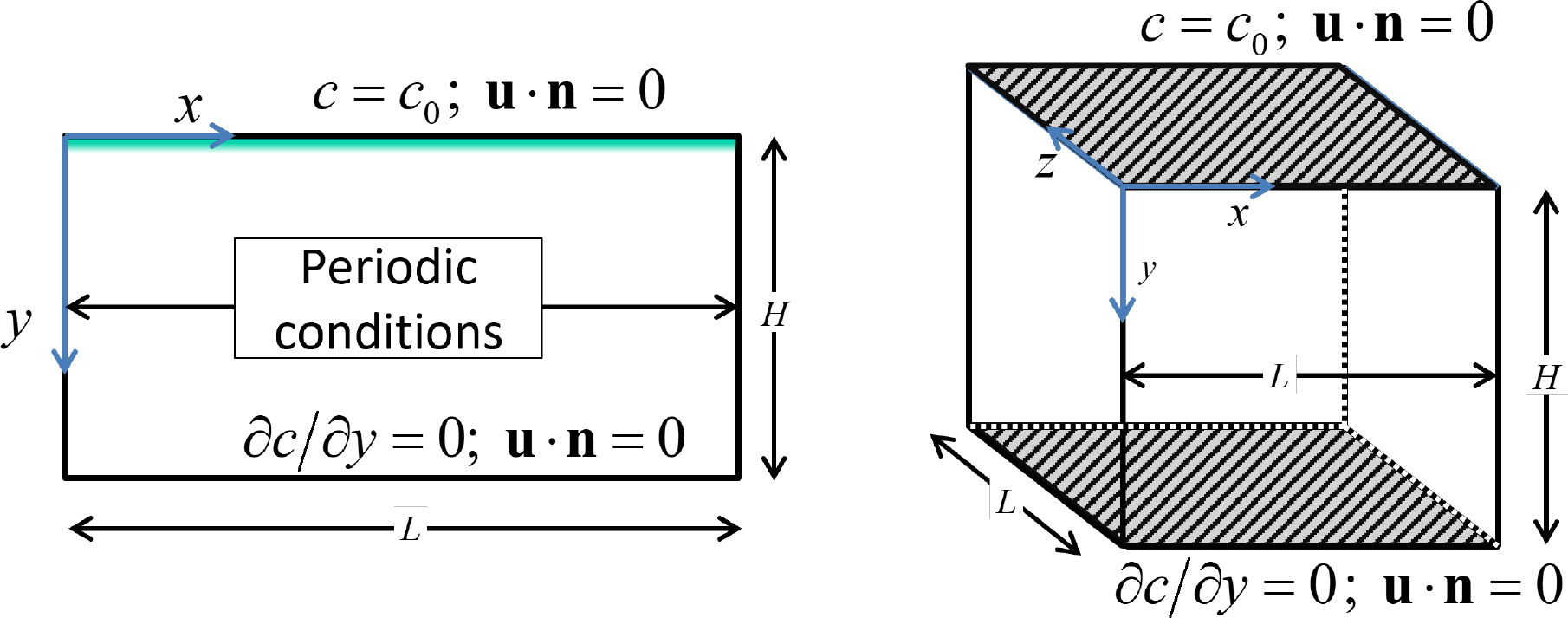}
\caption{\label{fig:flow_cells} Schematics of the two-dimensional (a) and three-dimensional (b)  configurations and boundary conditions; the boundary conditions associated to the governing equations \eqref{eqn:one} are indicated.}
\end{figure}
We consider a porous medium described at the Darcy (i.e., continuum) scale as homogeneous, i.e. with uniform porosity $\phi$, permeability $\kappa$, and  dispersivity lengths  $\alpha_\text{L}$ and $\alpha_\text{T}$, where the notations $\text{L}$ and $\text{T}$ refer respectively to the longitudinal and transverse direction with respect to the local velocity vector. The flow domain is assumed to be a rectangle (in two dimensions -- 2D) or a rectangle cuboid (in three dimensions -- 3D) of height $H$ in the vertical ($y$) direction and length $L$ in the horizontal direction(s) ($x$  in 2D, $x$ and $z$ in 3D), see Fig.~\ref{fig:flow_cells}. 

This domain is initially filled with pure brine. At the top boundary of the domain the brine is in contact with supercritical CO$_2$ (sCO$_2$, outside of the flow domain of Fig.~\ref{fig:flow_cells}) which  partially dissolves into the brine, rendering the density of the brine-CO$_2$ mixture dependent on the local concentration $c$ of the dissolved CO$_2$ according to a linear dependence $\rho=\rho_{0}+\left(c / c_{0}\right) \Delta \rho$, where $\rho_0$, $c_0$ and $\Delta \rho$ denote respectively the density of pure brine, the maximum concentration of the dissolved CO$_2$, which is controlled by its solubility in the brine, and the density difference between the mixture at concentration $c_0$ and the pure brine. 

The dissolution of supercritical CO$_2$ into brine is usually a much quicker process than the onset of instability, so its impact on the concentration of the dissolved CO$_2$ at the brine-sCO$_2$ interface ($y=0$ plane) can be safely assumed to result in a constant boundary concentration $c_0$ on that plane. The bottom boundary of the domain ($y=H$ plane) is assumed impermeable while 
periodic boundary conditions are imposed on all laterall boundaries. 
With this picture in mind, the boundary condition for the concentration $c$ and the velocity field $\mathbf{u}$ of the solution are as follows:
\begin{equation}
\begin{aligned}
\mathbf{u}(x, 0, [z,] t) \cdot \mathbf{\hat{y}} & =0 ; &  c(x, 0, [z,], t)& =c_{0}\\
\mathbf{u}(x, H, [z,], t)  \cdot \mathbf{\hat{y}} & =0 ; & \frac{\partial c}{\partial y}(x, H, [z,], t)& =0\\
\mathbf{u}(0, y, [z,], t) & =\mathbf{u}(L, y, [z,], t) ; & c(0, y, [z,] t)& =c(L, y, [z,] t)\\
\big [ ~~
\mathbf{u}(x, y, 0, t) \big . & =\mathbf{u}(x, y, L, t);  & c(x, y, 0, t)& = \big . c(x, y, L, t) ~~
\big ]
\end{aligned}
\end{equation}
where $t$ is time, $\mathbf{\hat{y}} $ is the unit vector along axis $y$, and the brackets $[ \cdots ]$ indicate equations or terms that are only present for the 3D geometry.
 
The classical Boussinesq approximation being applied on account of the small CO$_2$ dissolution, the mass conservation and coupled flow and solute transport in the porous medium are described respectively by the following  governing equations:
\begin{equation} 
\label{eqn:one}
\begin{aligned}
&\boldsymbol \nabla \cdot \mathbf{u}=0\\
&\mathbf{u}=-\frac{\kappa}{\eta}\left (\boldsymbol \nabla p-\rho(c) g  \,\hat{\mathbf{y}} \right )\\
&\phi \frac{\partial c}{\partial t} + \mathbf{u} \cdot \boldsymbol \nabla c=\phi \boldsymbol \nabla \cdot \left (\overline{\mathbf{D}} \cdot \boldsymbol \nabla c \right )
\end{aligned}
\end{equation}
where $p$ is the pressure field and $\eta$ the fluid's viscosity (assumed independent of $c$), while $\overline{\mathbf{D}}$ denotes the dispersion tensor. This tensor incorporates molecular diffusion but is also dependent on the local velocity through a non-diagonal component that accounts for hydrodynamic dispersion and is, in general, anisotropic even for an isotropic porous medium. Therefore, in a local reference frame attached to the local streamline, $\overline{\mathbf{D}}$ takes the form $\overline{\mathbf{D}}=D_{0}\overline{\mathbf{I}}+\overline{\mathbf{D}}_\text{hyd}$, where $D_0$ is the molecular diffusion coefficient, $\overline{\mathbf{I}}$ is the identity matrix and the {\em hydrodynamic} (or \textit{mechanical}) dispersion tensor $\overline{\mathbf{D}}_\text{hyd}$ has the form
\begin{equation}
\overline{\mathbf{D}}_\text{hyd}=\frac{1}{\phi} \left[\begin{array}{ccc}
\alpha_\text{L}\|\mathbf{u}\| & 0 \\
0 & \alpha_\text{T}\|\mathbf{u}\|
\end{array}\right] ~.
\end{equation} 
After transforming the dispersion tensor $\overline{\mathbf{D}}$ back into the $(x,y,z)$ cartesian reference frame, it takes the well-known form \citep{hidalgo2009effect}
\begin{equation}
\label{eq:Dij}
    D_{i j} = (D_0  + \frac{\alpha_\text{T}}{\phi} \|\textbf{u}\|) \delta_{ij} +  \frac{\alpha_\text{L}-\alpha_T}{\phi} \frac{u_i u_j}{\|\textbf{u}\|}
\end{equation}
where $\delta_{ij}$ is the Kronecker delta function and $\| \mathbf{u} \|$ the Euclidean norm of the velocity. 

\subsection{Non-dimensionalisation of governing equations}

We proceed to non-dimensionalize the above formulation by considering the following scales for length, velocity, time, pressure and concentration, respectively: $H$, $u_\text{ref}=\kappa \Delta \rho g / \eta$, $t_{\mathrm{ref}}=\phi H / u_{\mathrm{ref}}$, $\Delta \rho g H$ and $c_0$. The dimensionless form of the governing equations thus reads
\begin{equation}
\label{eq:nondim_eqs}
\begin{aligned} 
&\mathbf{\tilde{\boldsymbol \nabla}} \cdot \mathbf{\tilde{u}}=\mathbf{0}\\
&\mathbf{\tilde{u}}=-( \mathbf{\tilde{\boldsymbol \nabla}} \tilde{p} - \tilde{c} \, \hat{\mathbf{y}})\\
&\frac{\partial \tilde{c}}{\partial \tilde{t}} +\mathbf{\tilde{u}} \cdot  \mathbf{\tilde{\boldsymbol \nabla}} \tilde{c}=\frac{1}{R a}  \mathbf{\tilde{\boldsymbol \nabla}} \cdot(\tilde{\overline{\mathbf{D}}} \cdot  \mathbf{\tilde{\boldsymbol \nabla}} \tilde{c}) ~,
\end{aligned}
\end{equation}
where the $\tilde{~}$ sign denotes a non-dimensional quantity,  
\begin{equation}
\label{eq:def_Ra}
Ra= \frac{\Delta \rho g \kappa H}{\phi \eta D_0}
\end{equation}
is the Rayleigh number, and the dimensionless form of the dispersion tensor recasts as
\begin{equation}
\label{eq:nondim_disp_accurate}
\tilde{\overline{\mathbf{D}}}=\left (1+S \alpha \|\mathbf{\tilde{u}}\| \right ) \mathbf{I}+S(1-\alpha) \frac{\tilde{u}_{i} \tilde{u}_{j}}{\|\mathbf{\tilde{u}}\|} ~.
\end{equation}
Here
\begin{equation}
\label{eq:def_S_and_alpha}
S=\frac{\alpha_\text{L}  u_ \text{ref}}{D_\text{0} \phi} \text{~ ~ and ~ ~} \alpha = \frac{\alpha_\text{T}}{\alpha_\text{L}}
\end{equation}
are respectively the dispersion's strength as compared to molecular diffusion and the dispersivity ratio, which is a measure of the medium's  dispersive anisotropy. Note that the reference time $t_\text{ref}$ and Rayleigh number are independent of the dispersion properties, namely, dispersion strength and dispersivity ratio. This is appropriate as the dispersivity lengths are a property of the geometrical structure of the porous medium while molecular diffusion is typically independent of the medium's structure for the range of pore sizes significantly larger than $10$ \textmu{}m, which are characteristic of such subsurface porous media. As a consequence, increasing the dispersion strength in the model corresponds to increasing the grain size in dimensional units. This choice of dimensionalization is different from the one used by Hidalgo et al. \citep{hidalgo2009effect} who used the total diffusion $D_0+\alpha_L u_\text{ref}$ instead of $D_0$ to define the Rayleigh number; this difference in non-dimensionalization scheme will be further discussed in section~\ref{sec:roleOfDispersion}.

\subsection{Initial condition}

The initial condition to be applied to such a model has been the topic of numerous debates regarding its feasibility and realistic implications. It was delineated that instability structures that are triggered by the propagation of numerical errors may not match experimental results \citep{schincariol1994generation}. Such an observation follows from the fact that any pore-scale flow will experience continuous perturbations \citep{tilton_2018} that will lead to an occurrence of instability much before that predicted by numerical simulations where perturbations originate from numerical errors. Other studies have shown that the time of introduction of perturbation has no effect on the nonlinear onset time provided the perturbations are introduced relatively early \citep{selim2007stability,liu1997numerical}. Here we follow the usual procedure for instability initiation in this type of numerical simulations  \citep{ennis2005role,riaz2006onset,elenius2012time,ghesmat2011impact,Cheng2012WRR,slim2014solutal} to ensure the validity of the above points.  We first consider the background diffusive form, which is invariant in transverse directions and reads as
\begin{equation}
\label{eq:diffusive_Cprof}
\tilde{c}_\text{b}(\tilde{y}, \tilde{t})=1-\frac{4}{\pi} \sum_{n=1}^{\infty} \frac{1}{2 n-1} \sin \left ((n-1 / 2) \pi \tilde{y} \right ) \exp \left (- (n-1 / 2)^{2} \pi^{2} \tilde{t} / Ra\right) ~.
\end{equation}
It is devoid of dispersion since at initial times the velocity remains negligibly low.  We then superimpose to this diffusive form a perturbation  in the form (in 2D)
\begin{equation}
\tilde{c}_\text{p}=\mathbb{R} \: \xi \exp(-\xi^2) \text{~ with ~} \xi=\tilde{y} \sqrt{\dfrac{Ra}{4\tilde{t}_\text{p}}} ~,
\end{equation}
where $\mathbb{R}=\mathbb{R}(x)$ is a uniformly distributed random number with mean $0.5$, which  is kept the same throughout all the simulations, and $\tilde{t}_\text{p}$ is the time at which the perturbations are introduced.  For the 3D scenario, the form of the perturbation is similar with $\mathbb{R}=\mathbb{R}(x,z)$. All simulations are performed with $\tilde{t}_\text{p}=0.01$, thereby ensuring that the nonlinear onset time is much larger than the time at which perturbations are introduced for the given $Ra$ used in the study. Therefore, the net initial condition reads $\tilde{c}_{0}=\tilde{c}_{b}+\epsilon \tilde{c}_{p}$ with $\epsilon$ the amplitude of the perturbation. Note that since two systems with same values for $\tilde{t}_\text{p} Ra$ and $\epsilon \sqrt{Ra}$ are equivalent \citep{tilton2014nonlinear}, we have kept the value of $\tilde{t}_\text{p}$ and $\epsilon$ (set to 0.01) constant throughout the study. 

\subsection{Stream function formulation} \label{sec:2DFormulation}

\subsubsection{2D geometry}

Applying a stream function formulation to Eq.~\eqref{eq:nondim_eqs} and \eqref{eq:nondim_disp_accurate} yields the following equations of the stream function $\psi$ \citep{riaz2006onset,tilton2014nonlinear}:
\begin{equation} 
\label{eqn:2Dnondim}
\begin{aligned}
&\nabla^2\psi=\frac{\partial \tilde{c}}{\partial \tilde{x}}\\
&\frac{\partial \tilde{c}}{\partial \tilde{t}} +\mathbf{\tilde{u}} \cdot \mathbf{\tilde{\boldsymbol \nabla}} \tilde{c}=\frac{1}{Ra} \mathbf{\tilde{\boldsymbol \nabla}} \cdot(\tilde{\overline{\mathbf{D}}} \cdot \mathbf{\tilde{\boldsymbol \nabla}} \tilde{c}),
\end{aligned}
\end{equation}
where the dependence of $\tilde{\overline{\mathbf{D}}}$ on the velocity is given by Eq.~\eqref{eq:nondim_disp_accurate}. The velocities are derived from the stream function  according to
\begin{equation}
\tilde{u}_x=-\frac{\partial \psi}{\partial \tilde{y}}, \ \tilde{u}_y=\frac{\partial \psi}{\partial \tilde{x}} ~,
\end{equation}
and thus automatically satisfy the continuity equation. The stream function $\psi$ can hold arbitrary (gauge) values at the top and bottom walls, whereas its values are considered periodic at all lateral (vertical) boundaries. The boundary conditions for $\tilde{c}$ are the following:
\begin{equation}
\label{eq:BC_sf_2D}
\tilde{c}(\tilde{x}, 0,  \tilde{t}) = 1 \text{ ~, } \frac{\partial \tilde{c}}{\partial \tilde{y}}(\tilde{x}, 1, \tilde{t}) =0  \text{~ and ~}
\tilde{c} \left (0, \tilde{y}, \tilde{t} \right ) =\tilde{c}  \left (\frac{L}{H}, \tilde{y}, \tilde{t} \right )  ~.
\end{equation}

\subsubsection{3D geometry} \label{sec:3DFormulation}
Similarly, a stream formulation can be applied to Eq.~\eqref{eq:nondim_eqs} and \eqref{eq:nondim_disp_accurate} in a 3D flow domain \citep{Davis1989Stream3D}. The velocity field is inferred from the stream function through
\begin{equation}
\tilde{u}_x=\frac{\partial \psi}{\partial \tilde{y}},\ \tilde{u}_y=-\frac{\partial \psi}{\partial \tilde{x}}-\frac{\partial \theta}{\partial \tilde{z}},\ \ \tilde{u}_z=\frac{\partial \theta}{\partial \tilde{y}} ~,
\end{equation}
and thus automatically satisfies the continuity equation. The stream function is found by solving the following equations:
\begin{equation} 
\label{eqn:3Dnondim}
\begin{aligned}
&\nabla^2\theta-\frac{\partial ^2 \theta}{\partial \tilde{x}^2}+\frac{\partial ^2 \psi}{\partial \tilde{x} \partial \tilde{z}}=-\frac{\partial \tilde{c}}{\partial \tilde{z}}\\
&\nabla^2\psi-\frac{\partial ^2 \psi}{\partial \tilde{z}^2}+\frac{\partial ^2 \theta}{\partial \tilde{x} \partial \tilde{z}}=-\frac{\partial \tilde{c}}{\partial \tilde{x}}\\
&\frac{\partial \tilde{c}}{\partial \tilde{t}}=\mathbf{\mathbf{\tilde{u}}} \cdot \mathbf{\tilde{\boldsymbol \nabla}} \tilde{c}=\frac{1}{Ra} \mathbf{\tilde{\boldsymbol \nabla}} \cdot(\tilde{\overline{\mathbf{D}}} \cdot \mathbf{\tilde{\boldsymbol \nabla}} \tilde{c})
\end{aligned}
\end{equation}
where again the dependence of $\tilde{\overline{\mathbf{D}}}$ on the velocity field is given by Eq.~\eqref{eq:nondim_disp_accurate}. The stream functions $\theta$ and $\psi$ may hold independently arbitrary values at the top and bottom walls, while they are periodic at all the four lateral sides. The boundary conditions for $\tilde{c}$ are
\begin{equation}
\label{eq:BC_sf_3D}
\begin{aligned}
\tilde{c}(\tilde{x}, 0, \tilde{z}, \tilde{t}) = 1 & \text{ ~, } \frac{\partial \tilde{c}}{\partial \tilde{y}}(\tilde{x}, 1, \tilde{z}, \tilde{t}) =0 ~ ,\\
\tilde{c} \left (0, \tilde{y}, \tilde{z}, \tilde{t} \right ) =\tilde{c}\left (\frac{L}{H}, \tilde{y}, \tilde{z}, \tilde{t} \right)  & \text{ ~and~ } \tilde{c} \left (\tilde{x}, \tilde{y}, 0, \tilde{t} \right ) =  \tilde{c}\left (\tilde{x}, \tilde{y}, \frac{L}{H}, \tilde{t} \right ) ~.
\end{aligned}
\end{equation}

\subsection{Numerical Simulation}

For the two-dimensional flow domain, we solve the equation set~\eqref{eqn:2Dnondim} along with the boundary conditions \eqref{eq:BC_sf_2D}. In the three-dimensional domain, the set of governing equations solved are \ref{eqn:3Dnondim}, with boundary conditions \eqref{eq:BC_sf_3D}. The equations are solved numerically using the classical fine-volume discretization within the open source CFD toolbox \textit{OpenFOAM}. A Gauss bounded upwind scheme, which is a bounded scheme \citep{Warming1976Upwind}, is used for the discretization of the divergence terms, while the Laplacian terms are discretized using a Gauss linear corrected scheme, both of which are second-order accurate. The latter uses Gauss theorem to convert the volume integral to surface integral and then compute the surface fluxes using the given interpolating schemes. 
To maintain order consistency, the time is discretized using the classical bounded backward implicit scheme. With these numerical schemes, we have developed a fully two-way coupled Darcy-Transport equation solver within \textit{OpenFOAM} for both 2D and 3D cases wherein the  stream function equations are solved using the Preconditioned conjugate gradient (PCG) solver with Diagonal-based Incomplete Cholesky (DIC) preconditioner. The transport equation is solved using the stabilized Preconditioned bi-conjugate gradient with Diagonal-based Incomplete LU (DILU) preconditioner. A convergence tolerance of 10$^{-6}$ is used all throughout the simulation. For the 2D flow domain, a rectangular domain with $L=3 H$ and a grid size $1000\times1000$ are used.  For the 3D simulations, a cubic domain ($L = H$) with grid size $250\times250\times400$ for $Ra=1000$ and $350\times350\times600$ for $Ra=3000$ is used. The initial dimensionless $\Delta t$ is chosen as 10$^{-7}$ and a gradual time-adaptative scheme with maximum $\Delta t=0.01$ is used  based on the iterations needed to converge a particular set of coupled equations, thereby speeding up the overall solution process. For all the simulations, the domain is decomposed into smaller sections and the whole solver is parallelized across 8 logical cores for faster implementation.

The model and numerical codes were validated by comparison to data from \citep{tilton2013initial}, as explained in Appendix~\ref{appB}.

\subsection{Scalar quantities of interest}

From the simulated concentration and velocity fields, we analyze the impact of the strength of dispersion $S$, its anisotropy $\alpha$, and the Rayleigh number $Ra$ in terms of five  scalar observables. Firstly, the finger number density, i.e., the number of fingers within a transverse linear unit length  can be estimated at any time using from iso-$\tilde{c}$ plots, counting the number of region of maxima encountered (see for example in 3D Fig.~\ref{fig:Fingers3D}).

Secondly, the nonlinear onset time of the instability, i.e. the time at which the temporal evolution of the total vertical solute
flux starts to be impacted by the existence of convection velocities, thus deviating from the decreasing behavior attached to the purely-diffusive concentration profile \eqref{eq:diffusive_Cprof} and exhibiting a first trough that marks the onset time of the instability.  
At the top boundary of the flow domain, the velocities are horizontal and therefore the vertical solute flux is purely diffusive (no advective flux), but it is still impacted by convective velocities due to the dependence of the dispersion tensor on them. 
The dimensionless form of the flux reads in 2D as
\begin{equation}
\label{eq:flux}
\bar{J}(\tilde{t})=\frac{1}{Ra} \int_{\tilde{x}=0}^{L/H} \left (1+S \left \| \mathbf{\tilde{u}}(\tilde{x},0,\tilde{t}) \right  \|  \right)  \left . \frac{\partial \tilde{c}}{\partial \tilde{y}} \right )_{\tilde{y}=0} \mathrm{d}\tilde{x} ~,
\end{equation}
the 3D formulation involving a second integration over $\bar{z}$ between $0$ and $L/H$.

Thirdly, we compute the average CO$_2$ concentration in the flow domain, which in 2D is obtained as
\begin{equation}
\bar{c}(\tilde{t})= \frac{1}{A_{\Omega}}  \int_{\Omega}  \tilde{c}(\mathbf{\tilde{\textbf{x}}},\tilde{t}) \:\mathrm{d}\mathbf{\tilde{\textbf{x}}} ~,
\end{equation}
where $\Omega$ is the total flow domain, and $A_{\Omega}$ is its area (in 2D, or volume in 3D). 

Finally, we compute a measure of the mixing capacity within the liquid phase in the form of the scalar dissipation rate, the dimensionless form of which is \citep{Nicholas2013JCH,Marco2019PRFR}
\begin{equation}
\label{eq:scalardissipation}
\chi(\tilde{t})= \frac{1}{A_{\Omega}} \int_{\Omega} \left ( \tilde{\overline{\mathbf{D}}} \cdot \mathbf{\tilde{\boldsymbol \nabla}} \tilde{c}(\textbf{x},t) \right ) \cdot  \mathbf{\tilde{\boldsymbol \nabla}}  \tilde{c}(\mathbf{\tilde{\textbf{x}}},\tilde{t}) \: \mathrm{d}\mathbf{\tilde{\textbf{x}}} ~.
\end{equation}
The scalar dissipation rate is expected to approach zero at infinite times.


\section{Results}
\label{sec:results}

Here we present the impact the medium dispersion has on the convective dissolution dynamic in terms of the flow phenomenology (concentration and velocity fields), dissolution flux and nonlinear onset time, mean CO$_2$ concentration, and mixing strength.

\subsection{Phenomenology of the flow}
\label{subsec:phenomenology}

\begin{figure}
\centering
\includegraphics[width=0.99\textwidth]{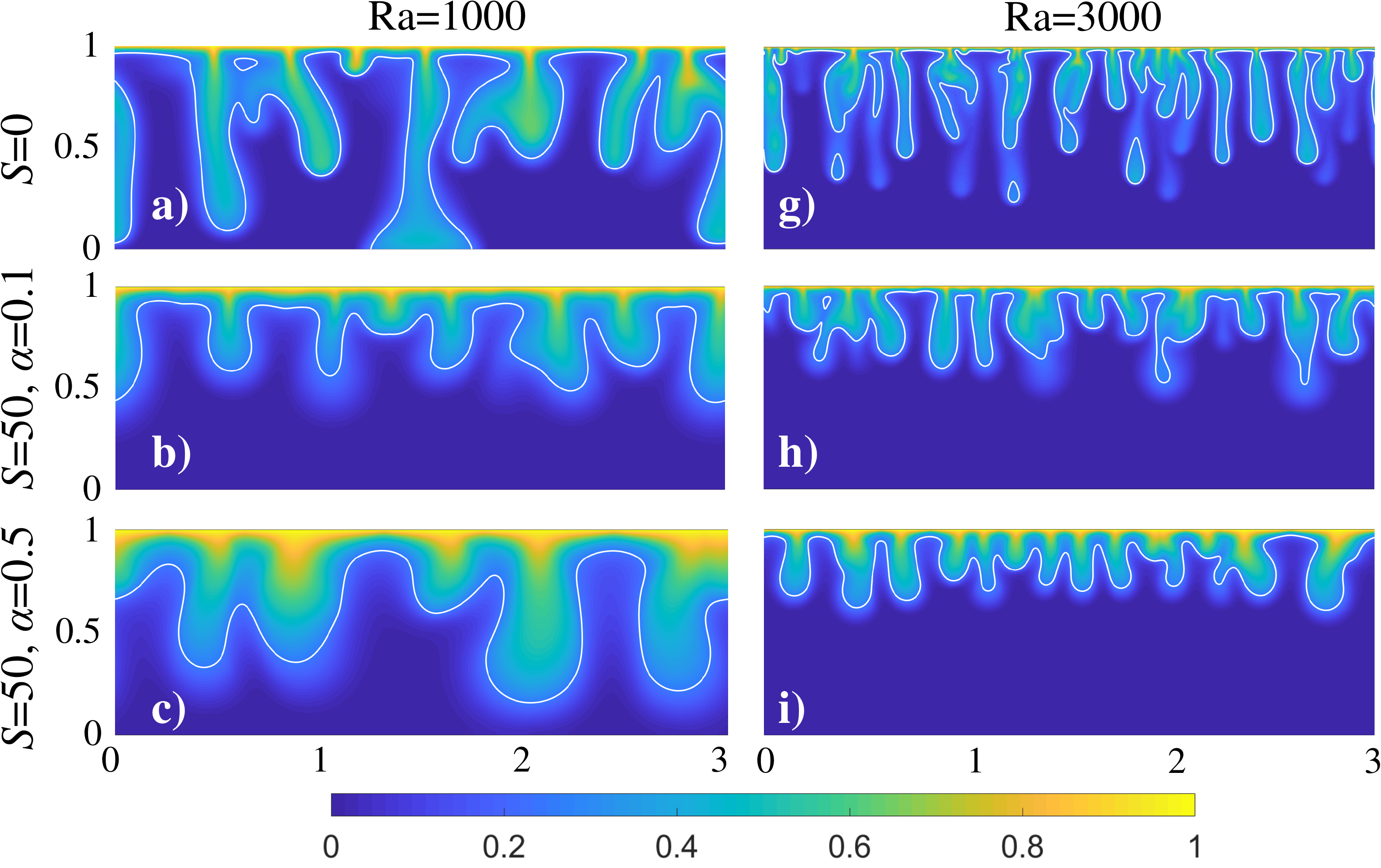}\\
\vspace*{1mm}
\includegraphics[width=0.99\textwidth]{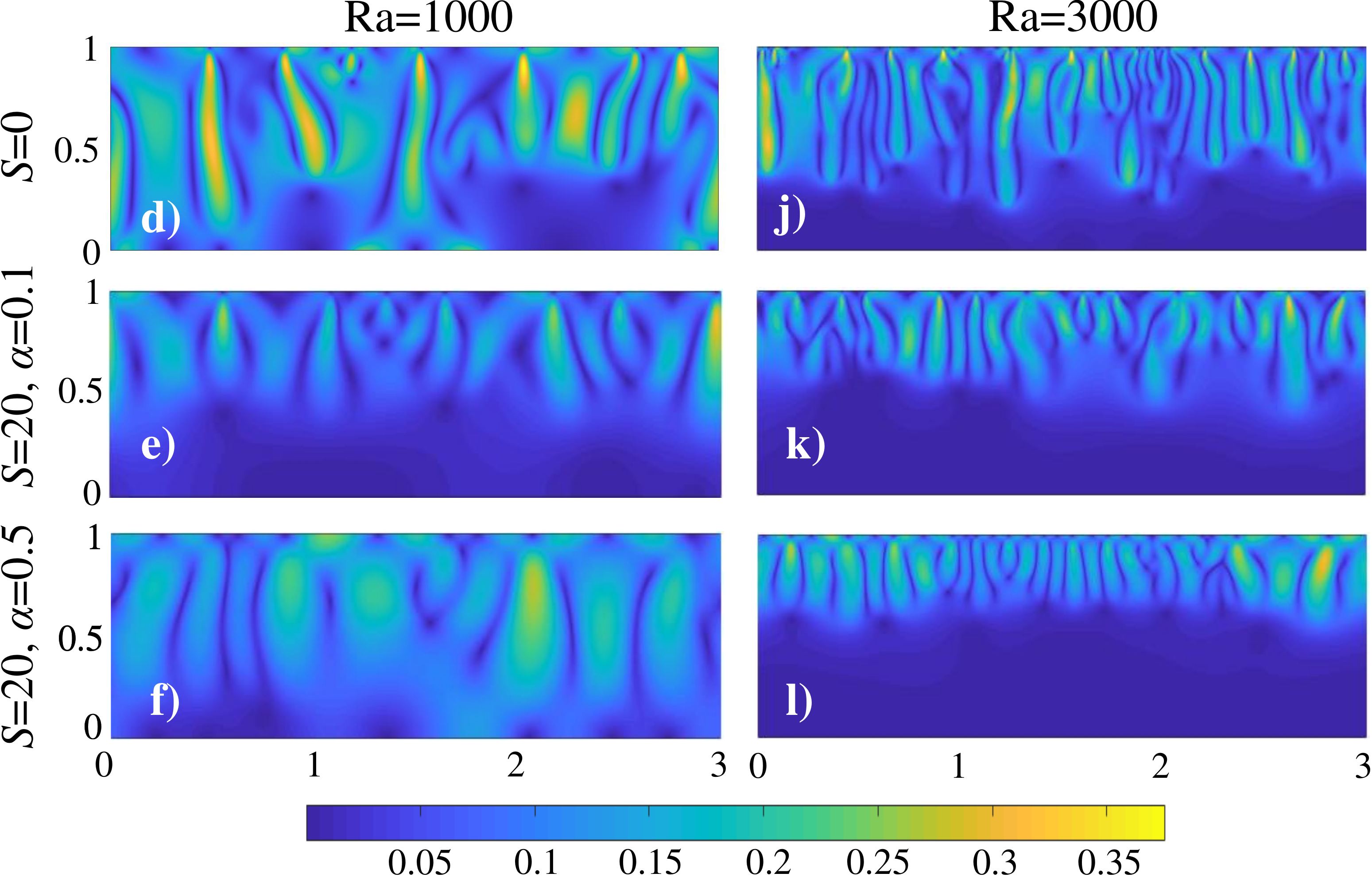}
\caption{\label{fig:conc_vel_profile} Comparison of concentration (top panel) and velocity (bottom panel) fields in the 2D geometry. In each panel the left column (a-f) shows the data at dimensionless time $\bar{t}=10$ for $Ra=1000$, while the right column (g-l) shows the data at dimensionless time $\bar{t}=6$ for $Ra=3000$. The first line of subfigures in each panel (a, d, g, j) corresponds to no hydrodynamic dispersion ($S=0$), the second one (b, e, h, k) to a strong but significantly anisotropic hydrodynamic dispersion ($S=50,\ \alpha=0.1$), and the third one (c, f, i, l) to a strong but isotropic dispersion ($S=50,\ \alpha=0.5$). The white line superimposed to the concentration maps is the iso-$\tilde{c}$ line at $\tilde{c}=0.25$.}
\end{figure}
Figure \ref{fig:conc_vel_profile} shows the concentration (left column) and velocity (right column) fields for two values of the Rayleigh number, $Ra=1000$ and $Ra=3000$ at a given time ($\tilde{t}=10$ and $6$, respectively), and for various configurations of the dispersion tensor. As reported in numerous previous experimental and numerical studies, the finger number density (FND), or wavenumber, is observed to increase with the Rayleigh number, i.e. between Fig.~\ref{fig:conc_vel_profile}a and \ref{fig:conc_vel_profile}g, \ref{fig:conc_vel_profile}b and \ref{fig:conc_vel_profile}h, as well as \ref{fig:conc_vel_profile}c and \ref{fig:conc_vel_profile}i.  When comparing the case of no mechanical dispersion ($S=0$) to that with a strong dispersion ($S=50$) and dispersivity ratio $\alpha=0.1$ typical of many sedimentary subsurface formations, the FND of the gravitational fingers is smaller for a larger dispersion strength (see Fig.~\ref{fig:conc_vel_profile}g and \ref{fig:conc_vel_profile}h), and the same holds for the penetration length of the fingers; this behavior is more pronounced for $Ra=3000$ than for $Ra=1000$. This is consistent with the stronger mixing associated with the larger effective diffusion coefficient resulting from mechanical dispersion. 
However, for a higher dispersivity ratio ($\alpha=0.5$), i.e.  a stronger transverse dispersivity, the finger penetration is larger than for $\alpha=0.1$ (see Fig.~\ref{fig:conc_vel_profile}c vs. \ref{fig:conc_vel_profile}b and \ref{fig:conc_vel_profile}i vs. \ref{fig:conc_vel_profile}h), while the finger density number is not much impacted. But the fingers are still  less advanced towards the bottom of the flow domain when mechanical dispersion is present, whatever the dispersivity ratio may be, as seen when comparing Fig.\ref{fig:conc_vel_profile}a with \ref{fig:conc_vel_profile}c or Fig.~\ref{fig:conc_vel_profile}g with Fig.~\ref{fig:conc_vel_profile}i.

\begin{figure}
\centering
\includegraphics[width=0.95\textwidth]{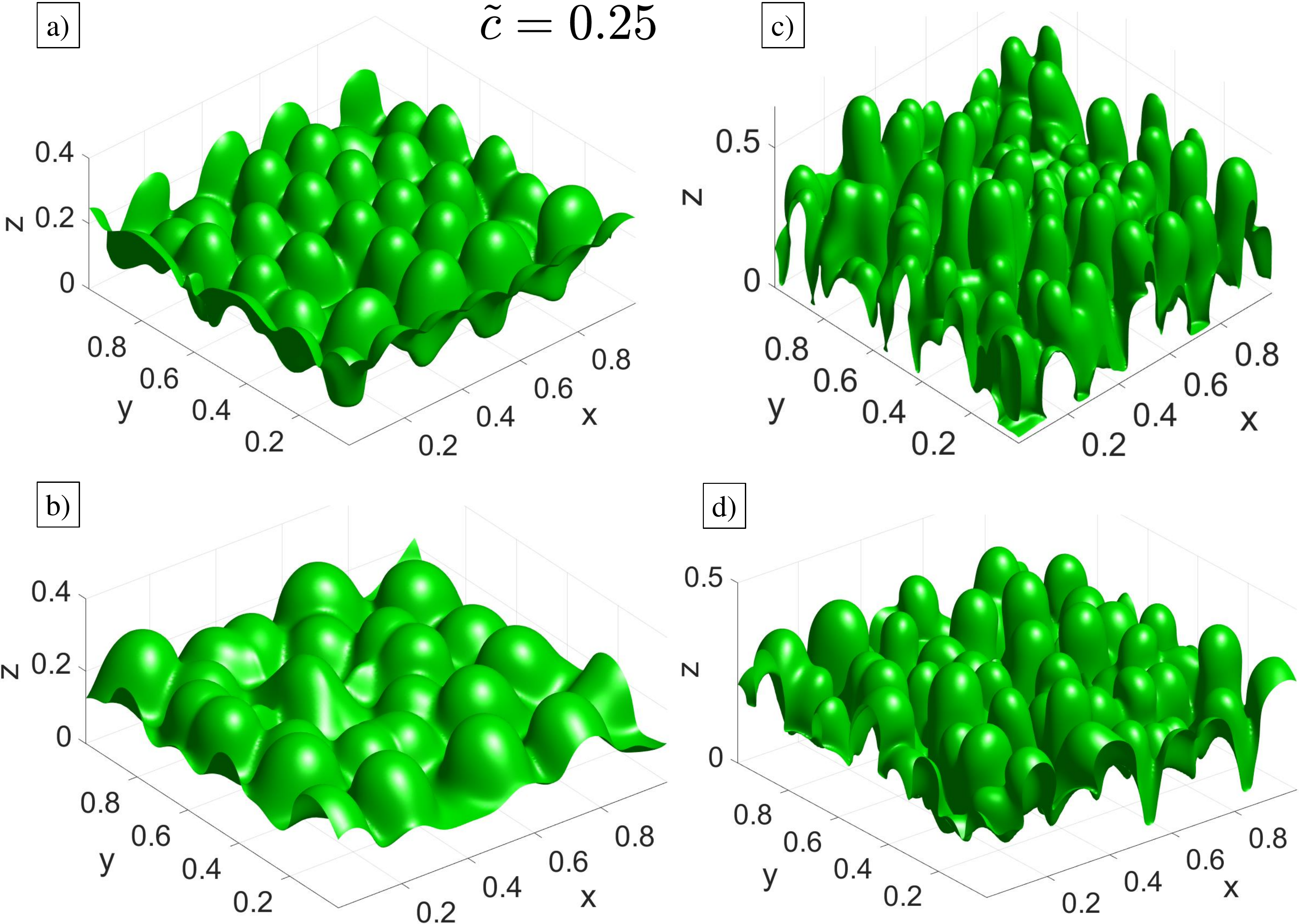}
\caption{\label{fig:Fingers3D}Comparison of inverted 3D finger patterns (i.e., iso-$\tilde{c}$ surfaces at $\tilde{c}=0.25$)  at dimensionless time $\bar{t}=4$, for $Ra=1000$ (left column) and $Ra=3000$ (right column), under different dispersion configurations: (a, c) $S=0$ and (b, d) $S=20$ with  $\alpha=0.1$.}
\end{figure}
This behavior is also observed on iso-surfaces $\tilde{c}=0.25$ of the 3D concentration,  as depicted at time $\bar{t}=6$ in Fig.~\ref{fig:Fingers3D}  for dispersion strengths $S=0$ and $S=20$. The finger number density is all the larger as the Rayleigh number is larger, as expected. Furthermore, the Rayleigh-Taylor instability exhibits thicker and more round-shaped  fingers when dispersion is significant, as compared to the case without hydrodynamic dispersion. As also observed in the 2D geometry (Fig.~\ref{fig:conc_vel_profile}), with a larger dispersion the finger penetration depth within the system is smaller. 

\begin{figure}
\centering
\includegraphics[width=0.95\textwidth]{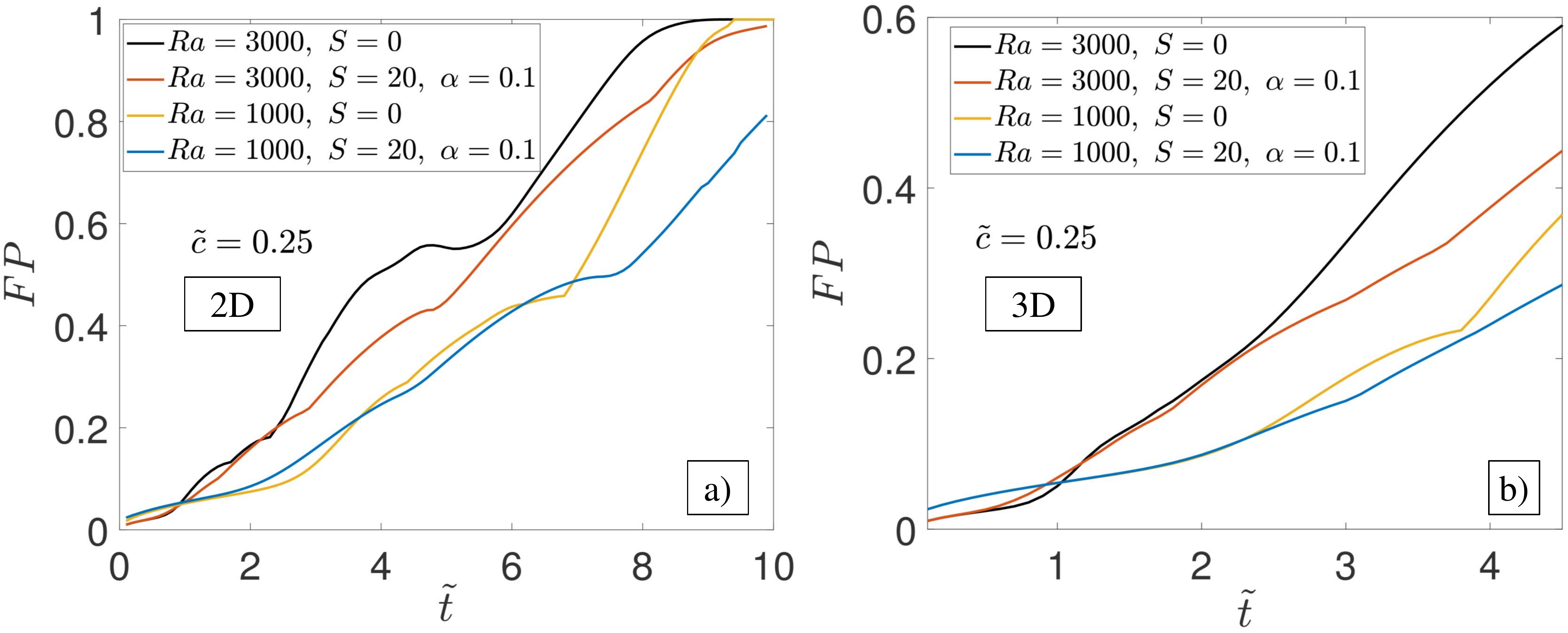}
\caption{\label{fig:penetration} Temporal evolution of the finger penetration depth (estimated from the $\bar{c}=0.25$ iso-line/surface) in the 2D (left)  and 3D (right) geometries, for  two Rayleigh numbers ($Ra=1000$ and $Ra=3000$) and two dispersion configuration (no dispersion and $S=20$, $\alpha=0.1$).}
\end{figure}
The latter conclusion relative to the penetration depth at a given non-dimensional time is confirmed by  Fig.~\ref{fig:penetration}: increasing the Rayleigh leads to faster penetration, and so does hydrodynamic dispersion (in comparison to pure molecular diffusion) with a rather standard dispersivity ratio of $\alpha=0.1$.
\begin{figure}
\centering
\includegraphics[width=0.98\textwidth]{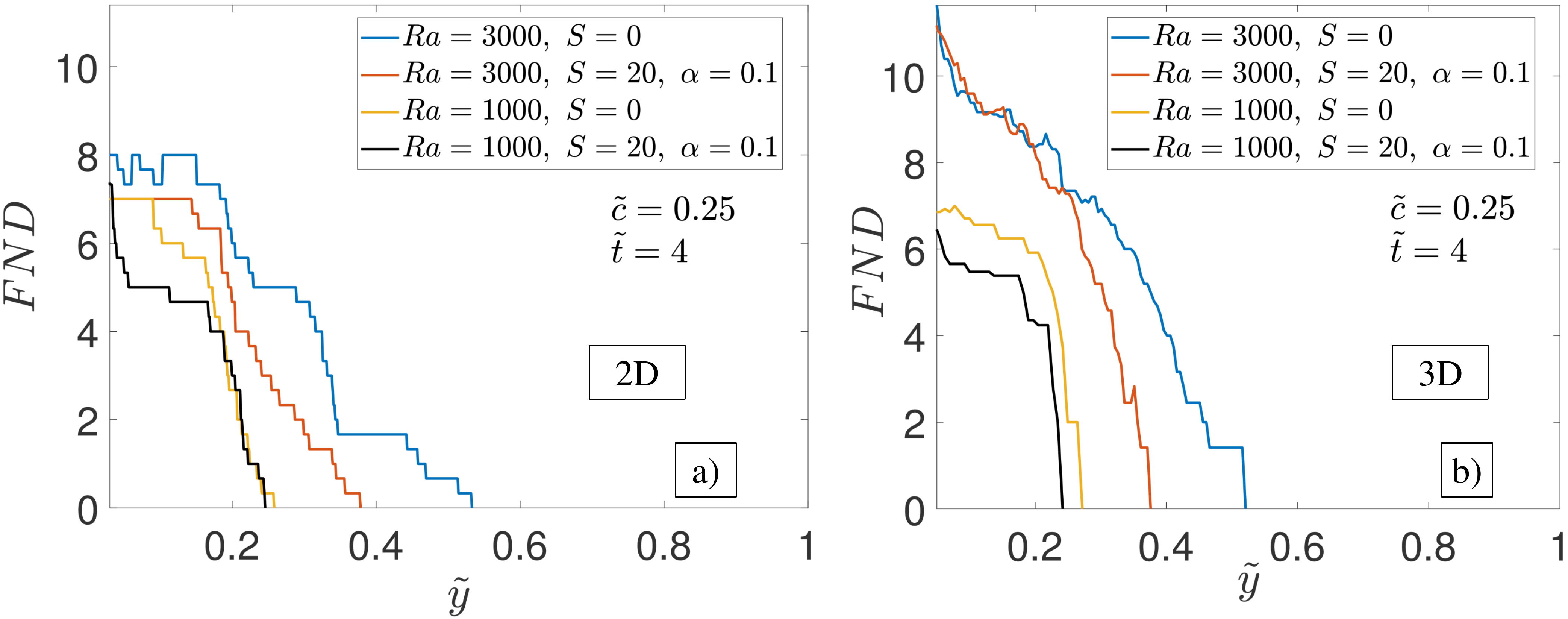}
\caption{\label{fig:FND_variation} 
Vertical profile of the finger number density in the 2D (left) and 3D  (right) geometries, for  two Rayleigh numbers ($Ra=1000$ and $Ra=3000$) and two dispersion configurations (no dispersion and $S=20$, $\alpha=0.1$).}
\end{figure}
Let us now examine vertical profiles of FND for various combinations of $Ra$ numbers and dispersion strengths $S$ (Figure~\ref{fig:FND_variation}). 
The figure confirms the qualitative observation of Figures~\ref{fig:conc_vel_profile} and \ref{fig:Fingers3D}. At a given time, the FND decreases with the vertical coordinate due to transverse coalescence of the solute fingers, which contributes to maintaining a high concentration of the solute in the fingers and thus driving the convection. Note that the vertical FND profile flattens in time (in Fig.~\ref{fig:FND_variation}) only one time is shown, as the large FND near the top boundary of the domain decreases while the penetration depth of the fingers increases (Fig.~\ref{fig:FND_variation}b and \ref{fig:FND_variation}d).  This has been previously observed in numerous experiments  (e.g, \citet{fernandez2002density,nakanishi2016experimental,wang2016three}). The FND is always larger at any vertical position for $Ra=3000$ than for $Ra=1000$ (Fig.~\ref{fig:FND_variation}a and \ref{fig:FND_variation}c). 
However, with an increase in hydrodynamic dispersion within the medium, increased transverse dispersion accelerates finger coalescence as concentration fingers progress towards the bottom of the flow domain. For a given $Ra$ number, an increase in dispersion $S$ thus results in a smaller finger number density (Fig.~\ref{fig:FND_variation}b and d). For $Ra=1000$ this difference is significant at small times but decreases with time, whereas for $Ra=3000$ the impact of dispersion (as compared to the pure molecular diffusion case) increases in time. These observations remains persistent across different values of the dispersivity ratio $\alpha$ (not shown in the figures).

\begin{figure}
\centering
\includegraphics[width=0.99\textwidth]{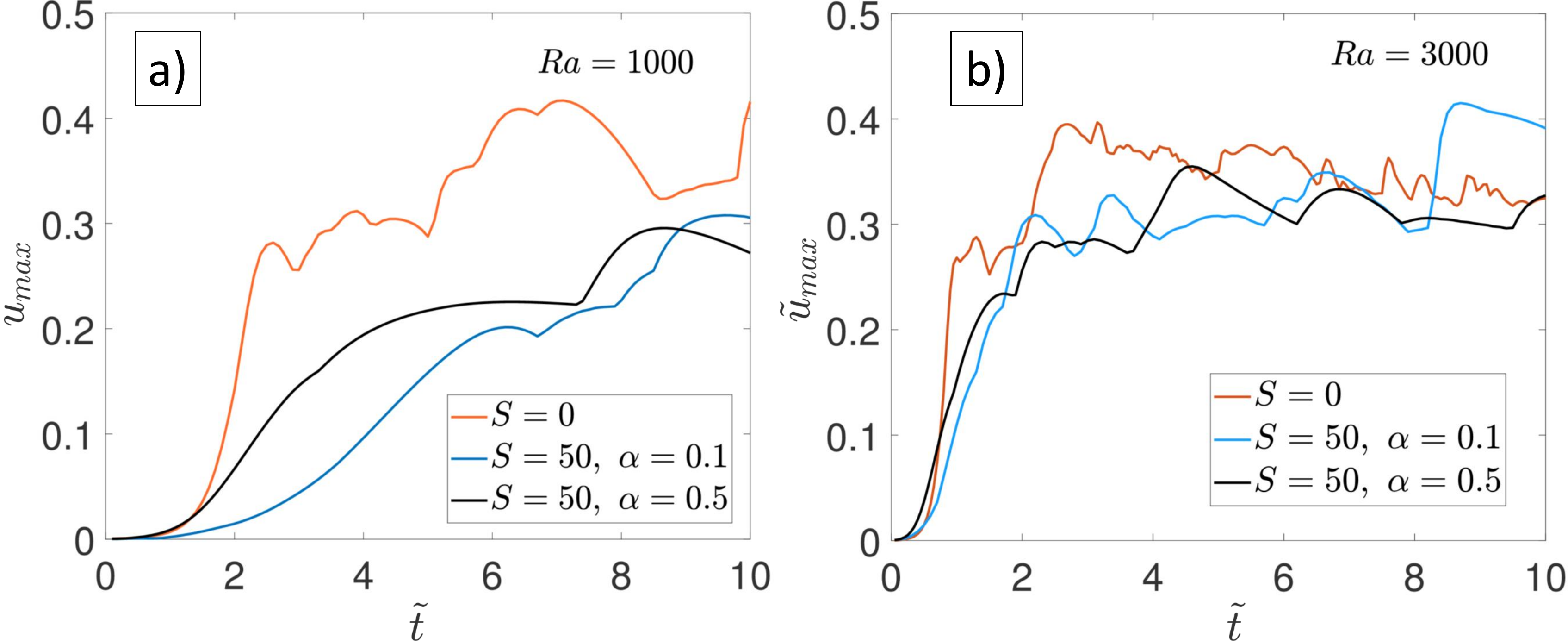}
\caption{\label{fig:velocity_profile} Temporal evolution of the maximum flow velocity in the 2D domain for  the dispersion configurations addressed in Fig.~\ref{fig:conc_vel_profile}: a) $Ra=1000$ and b) $Ra=3000$.}
\end{figure}
The velocity fields shown in the lower panel of Fig.~\ref{fig:conc_vel_profile} for the same flow conditions as the concentration fields shown in the upper panel of the same figure, reflect the convective strength of the instability process. The dependence of the mechanical dispersion on the local velocity magnitude introduces an additional coupling between velocities and concentrations: higher velocity magnitudes incur a higher impact of dispersion on the resulting dynamics. A general observation from the velocity maps in the lower panel of Fig.~\ref{fig:conc_vel_profile} is that the velocity magnitude near the top boundary is small but that a non-zero horizontal component exists, due to the slip condition imposed on velocity at the top boundary of the domain, which corresponds to the free surface between supercritical CO$_2$ and the aqueous phase (see the maps of $\tilde{v}_x$ in Fig.~\ref{fig:Vx_and_Vy}a in Appendix~\ref{sec:AppVxVz}).  This tangential velocity field at the wall also contributes to the net dispersive flux through Eq.~\eqref{eq:nondim_disp_accurate}. The velocity profile also exhibits alternating regions of high and low magnitudes, and the general structure reflects that of the concentration maps, with thinner fingers for a larger Rayleigh. The high magnitude regions are signatures of faster penetrating downward fingers, as well as of regions where the resident fluid comes up to replace the mixture near the top boundary of the domain. Unsurprisingly, boundaries between those regions of downward and upward flow exhibit a very low velocity. Comparing plots  (d), (e) and (f) in the lower panel of Figure~\ref{fig:conc_vel_profile}, we investigate the impact of hydrodynamic dispersion on the spatial distributions of the velocity magnitude. The slow-down of descending fingers by dispersion is obvious for $\alpha=0.1$ (Fig.~\ref{fig:conc_vel_profile}e), with thinner and shorter fingers (see also the $\tilde{v}_z$ maps in in Fig.~\ref{fig:Vx_and_Vy}a in Appendix~\ref{sec:AppVxVz})) and lower velocity magnitudes as compared to the configuration with no dispersion (Fig.~\ref{fig:conc_vel_profile}d).  In contrast, for $\alpha=0.5$ (Fig~\ref{fig:conc_vel_profile}f), the velocity maps show much broader fingers than for $\alpha=0.1$ (Fig~\ref{fig:conc_vel_profile}e), hereby covering a larger section of the domain with high velocity fields. 
The temporal evolution of the largest velocity in the 2D domain is shown in Fig.~\ref{fig:velocity_profile} for $Ra=1000$ and $Ra=3000$, and for three dispersion configurations. For $Ra=3000$, and for all dispersion configurations, the largest velocity reaches some sort of plateau around which it fluctuates, while for $Ra=1000$ we expect it to behave in the same way but the plateau is hardly reached at our maximal simulation time, $\tilde{t}=10$, for $S=50$.
As also reported previously (\citep{xie2011speed,emami2017dispersion}), this plateau of the finger penetration velocity does not vary drastically with dispersion. Hydrodynamic dispersion slows down the transitory regime before the plateau is reached, but the dispersivity ratio $\alpha$ seems to have little impact. Note that we do not show the corresponding plots obtained in the 3D domain, as the maximal investigated time is $\tilde{t}=4$ for the 3D data, which is a bit early to conclude on the maximal velocity behavior.

\subsection{Flux and onset time} \label{sec:FluxOnset}
\begin{figure}
\centering
\includegraphics[width=0.99\textwidth]{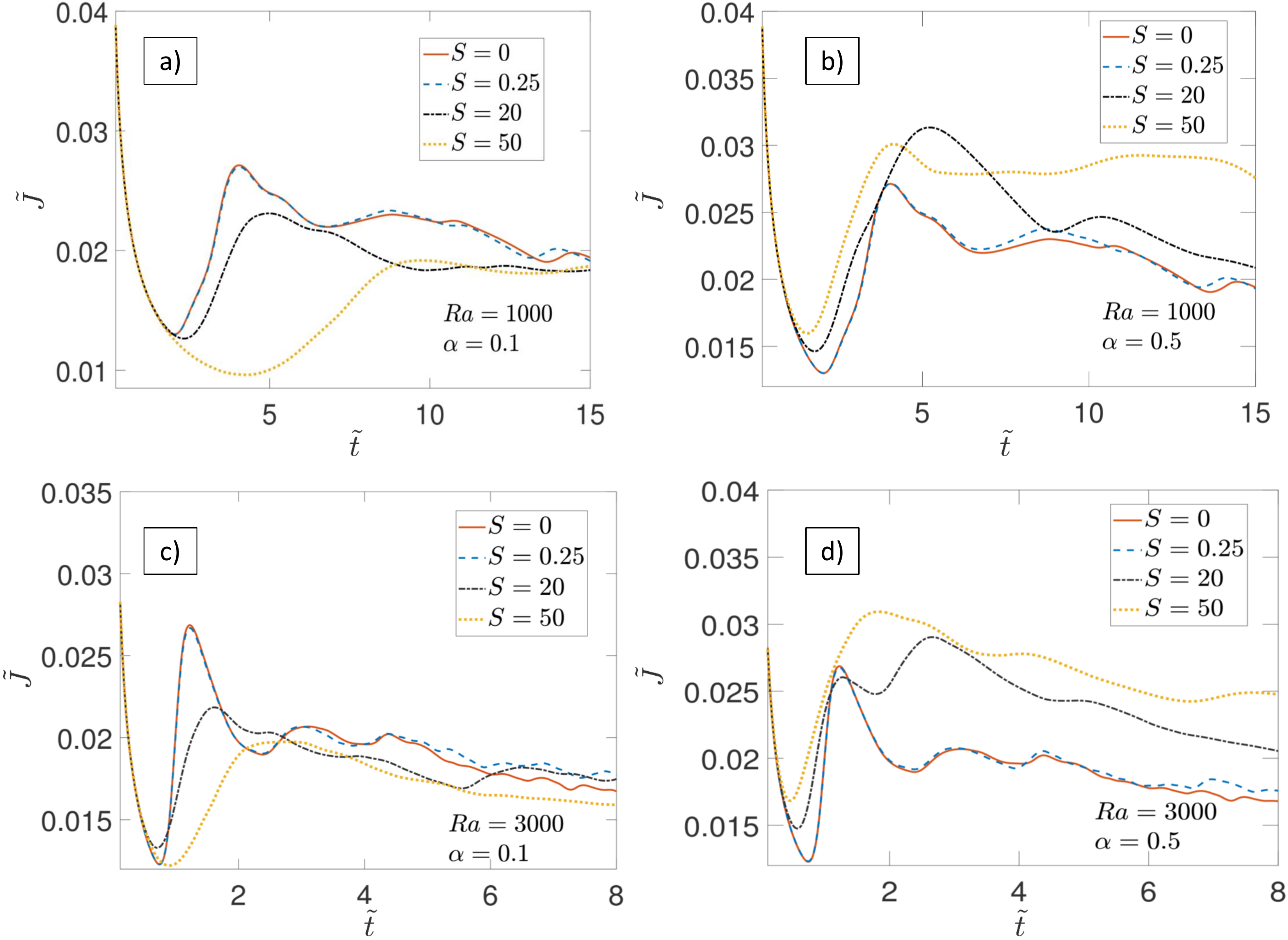}
\caption{\label{fig:Flux} Temporal evolution of the dispersive flux at the top boundary of the 2D flow domain for Rayleigh numbers $Ra=1000$  -- top line, i.e. (a, b) -- and $Ra=3000$  -- bottom line, i.e. (c, d) -- and for different values of the dispersive strength. The dispersivity ratio is  $\alpha=0.1$ -- left column, i.e. (a, c) --  or $\alpha=0.5$ -- right column, i.e. (b,d).}
\end{figure}

\begin{figure}
\centering
\includegraphics[width=0.99\textwidth]{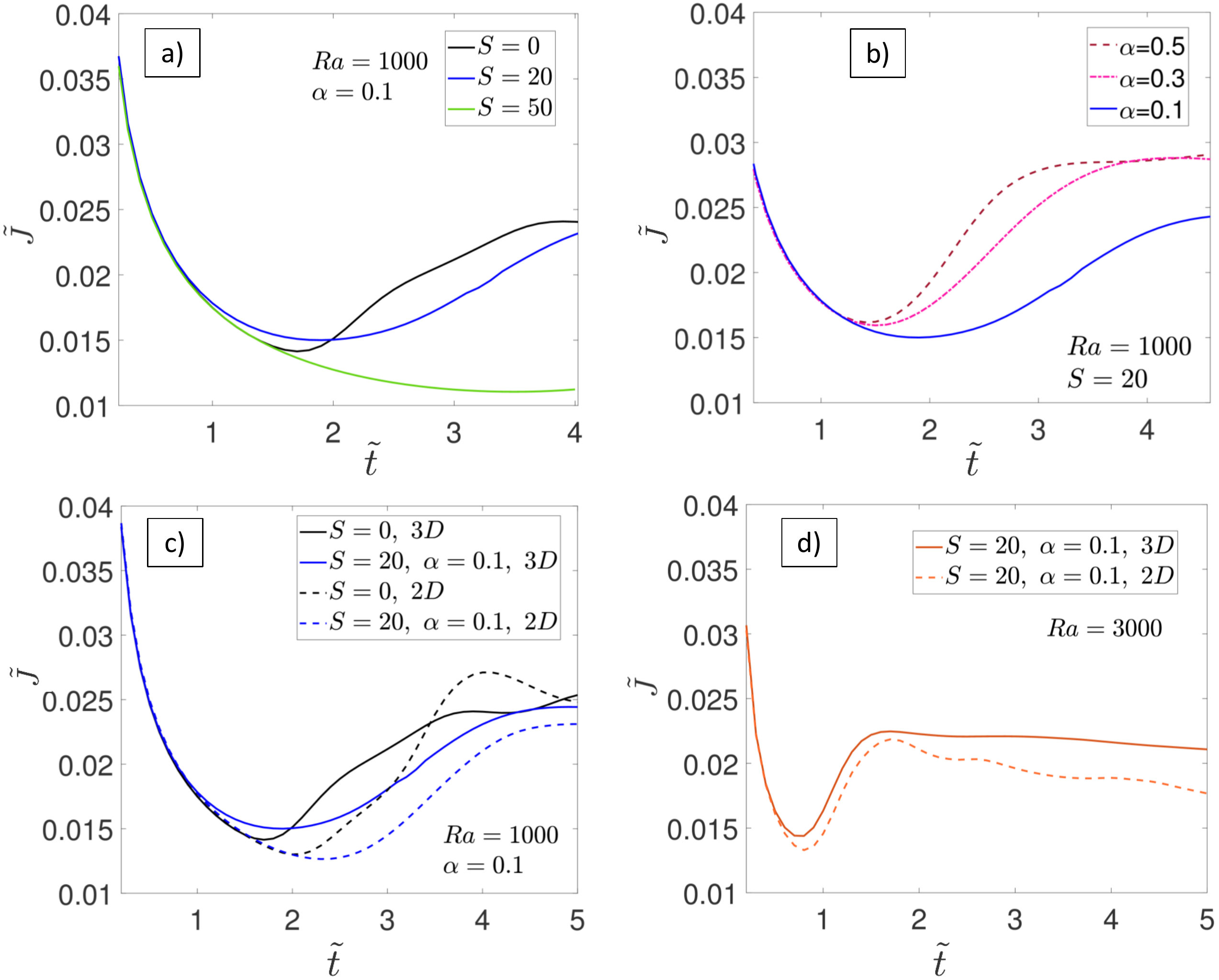}
\caption{\label{fig:compare2D3DFlux}(a,b): Temporal evolution of the dispersive flux at the top boundary of the 3D flow domain for Rayleigh number $Ra=1000$: (a) for dispersion strength $D=20$ and different values of the dispersive strength, $\alpha=0.1$, $0.2$ and $0.3$; (b) for dispersivity ratio $\alpha=0.1$ and different dispersion strengths $S=0$, $20$ and $50$. (c, d): Comparison between the results from the 2D and 3D simulations for $\alpha=0.1$ and two dispersion strengths $S=0$ and $20$: (c) $Ra=1000$ and (d) $Ra=3000$.}
\end{figure}

In light of these observations, the flow structure and concentration field are obviously strongly influenced by the strength of hydrodynamic dispersion as well as the medium's anisotropy. Therefore, we expect these quantities to impact the onset of convection and the dissolution rates in the medium. 

\begin{figure}
\centering
\includegraphics[width=0.95\textwidth]{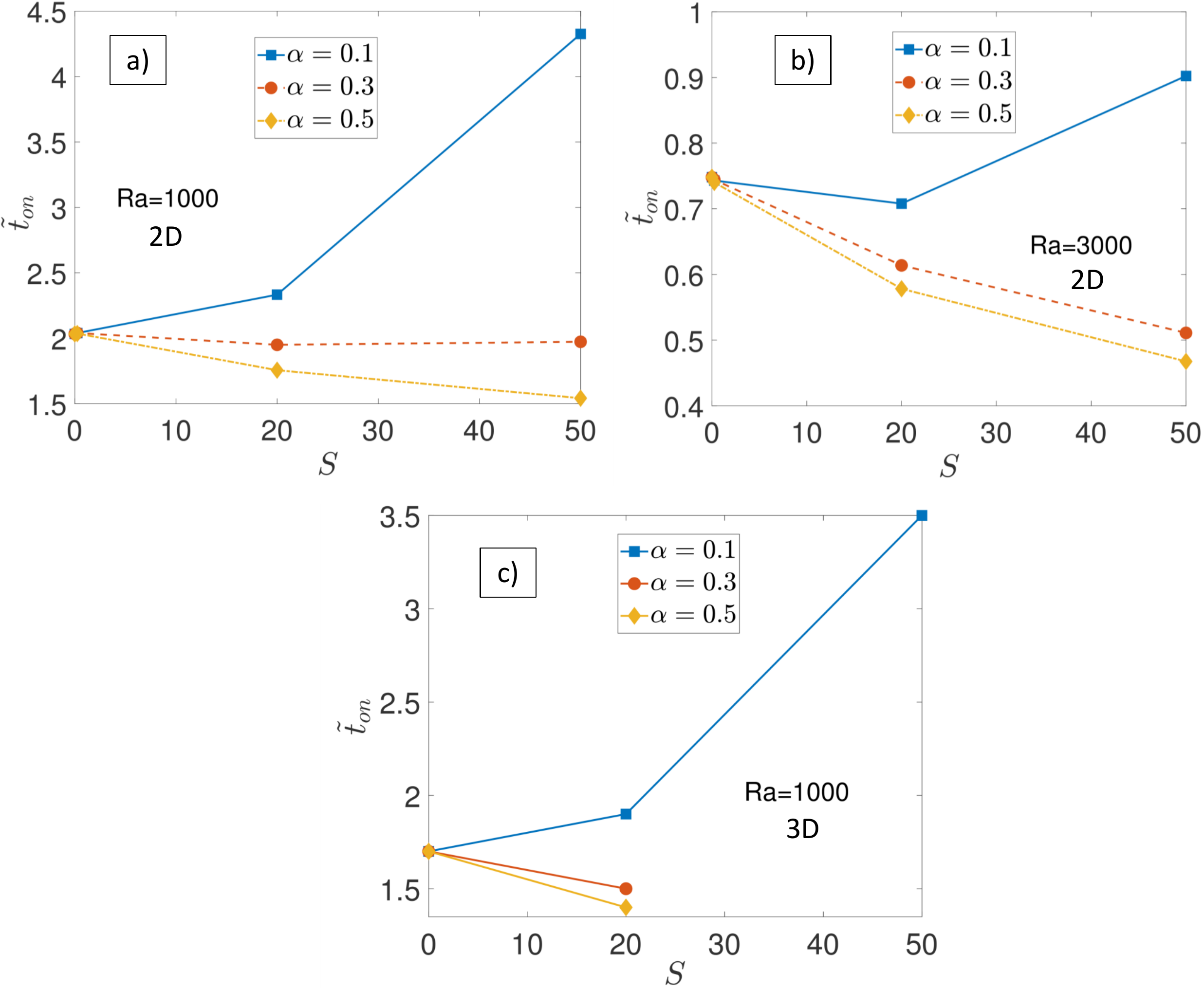}
\caption{\label{fig:OnsetTime} Dependence of onset time on mechanical dispersion strength $S$ for various values of the dispersivity ratio $\alpha$ (a) in the 2D geometry for (a) $Ra=1000$ and (b) $Ra=3000$, and (c) in the 3D geometry for $Ra=1000$.}
\end{figure}

Fig.~\ref{fig:Flux} shows the temporal evolution of the flux $\tilde{J}$ in the 2D geometry, for different Rayleigh numbers $Ra$ and dispersion strengths $S$, while Fig.~\ref{fig:compare2D3DFlux} shows similar data obtained in the 3D geometry. The flux profile exhibits an initial diffusive regime followed by a minimum which indicates the onset of convection. 
In contrast to what was reported previously by \citet{ghesmat2011effect},  the variation in onset time when varying the parameters  that control hydrodynamic dispersion, ($S$ and $\alpha$) are not drastic; but they are significant. We'll discuss in section~\ref{sec:roleOfDispersion} why our findings differ on this aspect from that of \citet{ghesmat2011effect} and other authors. 
Comparing the temporal evolution of the fluxes for $Ra=1000$ and $Ra=3000$ (i.e., Fig.~\ref{fig:Flux}a and \ref{fig:Flux}b to \ref{fig:Flux}c and \ref{fig:Flux}d, respectively), we oberve that an increase in Rayleigh number decreases the onset time, as already observed by several studies previously. Considering dispersivity ratios $\alpha=0.1$ and $\alpha=0.5$, both at $Ra=1000$ (Fig.~\ref{fig:Flux}b and Fig.~\ref{fig:Flux}d) and $Ra=3000$, and varying the dispersion strength between $S=0$ and $S=50$, we see that the onset time varies with $S$ but in a manner that depends both on $Ra$ and $\alpha$.  The results are summarized in Fig.~\ref{fig:OnsetTime}. For $Ra=1000$ in the 2D geometry (Fig.~\ref{fig:OnsetTime}a), an increase in mechanical dispersion $S$ delays the onset when $\alpha=0.1$, but for $\alpha=0.3$ and $\alpha=0.5$ (more so in the latter case), on the contrary, the onset time is decreased with increasing $S$. Furthermore, at any investigated $S$ value, increasing the dispersivity ratio decreases the onset time. For $Ra=3000$ in the 2D geometry (Fig.~\ref{fig:OnsetTime}b), a similar trend is observed when changing the dispersivity ratio at a given dispersion strength: the onset times decreases with $\alpha$; it also decreases monotonically with $S$ for $\alpha=0.3$ and $\alpha=0.5$, but for $\alpha=0.1$ the plot is not monotonic: it  decreases between $S=0$ and $S=20$, and increases between $S=20$ and $S=50$. The temporal evolution of the flux in 3D cases shows a behavior consistent with the 2D observations: for $Ra=1000$ and $\alpha=0.1$, increasing $S$ delays the onset of convection, while for $S=20$ increasing $\alpha$ leads to a smaller onset time. The results obtained in the 3D geometry are summarized in Fig.~\ref{fig:OnsetTime}c for $Ra=1000$; the dependence on dispersion parameters is fully consistent with the results obtained for the equivalent configuration in 2D (Fig.~\ref{fig:OnsetTime}a). Furthermore, when comparing the onset times obtained in the 2D and 3D geometry, all parameters being identical otherwise, the onset is observed to occur slightly earlier in the 3D geometry. This difference between onset times in the 2D and 3D geometries seems all the more apparent as the $Ra$ number is smaller (see Fig.~\ref{fig:compare2D3DFlux}a vs. Fig.~\ref{fig:compare2D3DFlux}b) . 

Besides the variation in the onset time, a large impact of hydrodynamic dispersion is observed after the convective instability kicks in. 
The overall flux in the constant flux regime, where the flux oscillates around a plateau that is more clearly defined, and over a larger time range, for higher Rayleigh numbers \citep{emami2015convective} (see Fig.~\ref{fig:Flux}c and \ref{fig:Flux}d as compared to \ref{fig:Flux}a and \ref{fig:Flux}b, respectively) tends to decrease slowly with time. Nevertheless, that plateau flux remains higher than the pure diffusive flux, and is slightly larger for $Ra=1000$ than for $Ra=3000$. It is also reached earlier for $Ra=3000$ than for $Ra=1000$, as the onset time is lower in the latter case.  The impact of the dispersion parameters on the plateau is qualitatively similar for the two investigated Rayleigh values. At a given Rayleigh number, increasing the dispersion's strength at $\alpha=0.1$ decreases the mean flux in the constant flux regime,  while for $\alpha=0.5$ the opposite behavior is observed.

\subsection{Mean concentration}

\begin{figure}
\centering
\includegraphics[width=0.95\textwidth]{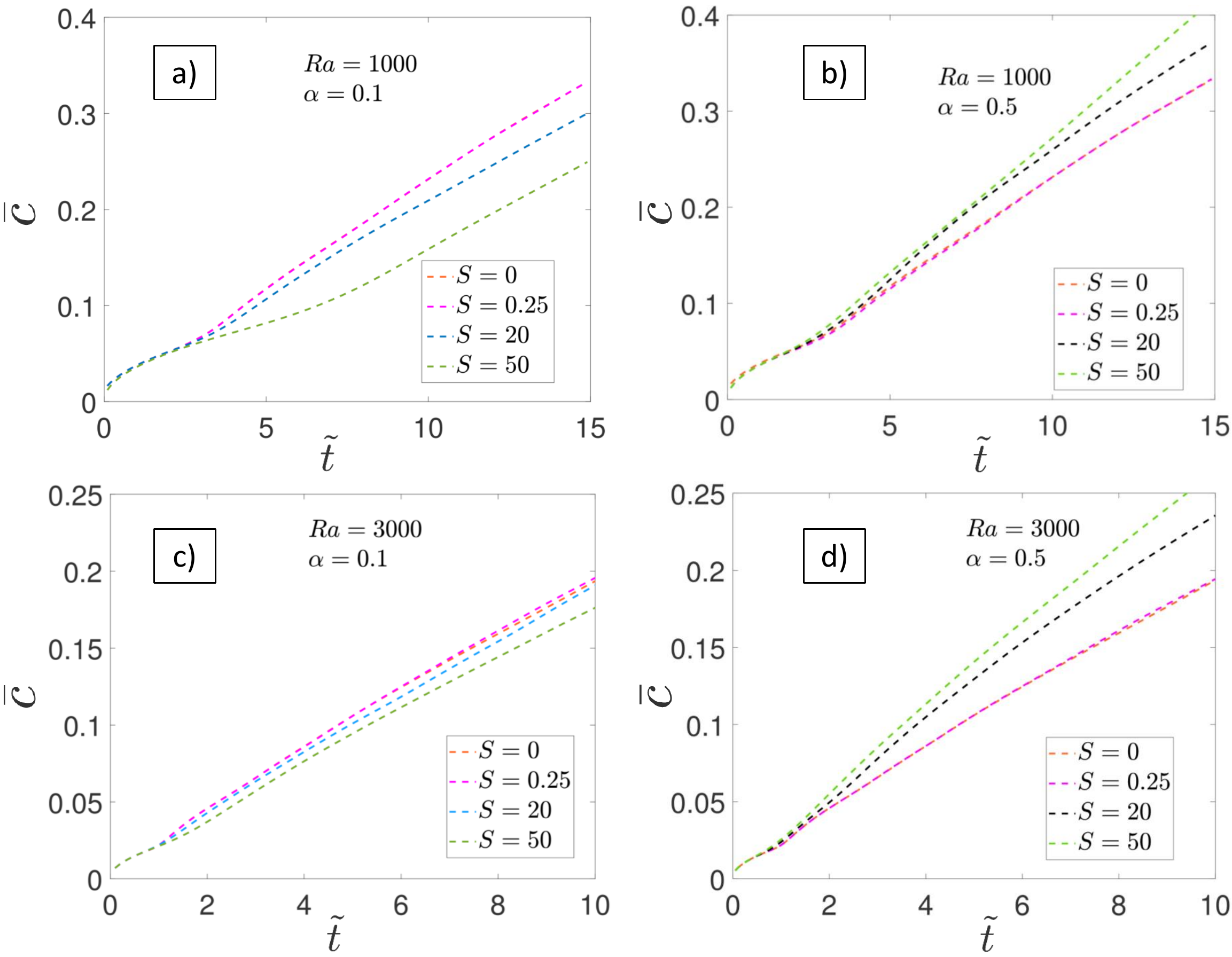}
\caption{\label{fig:Dissolution} Temporal evolution of the mean concentration in the entire flow domain for $Ra=1000$ (a, b) and $Ra=3000$ (c, d), and for $\alpha=0.1$ (a, c) and $\alpha=0.5$ (b, d).}
\end{figure}
\begin{figure}
\centering
\includegraphics[width=0.9\textwidth]{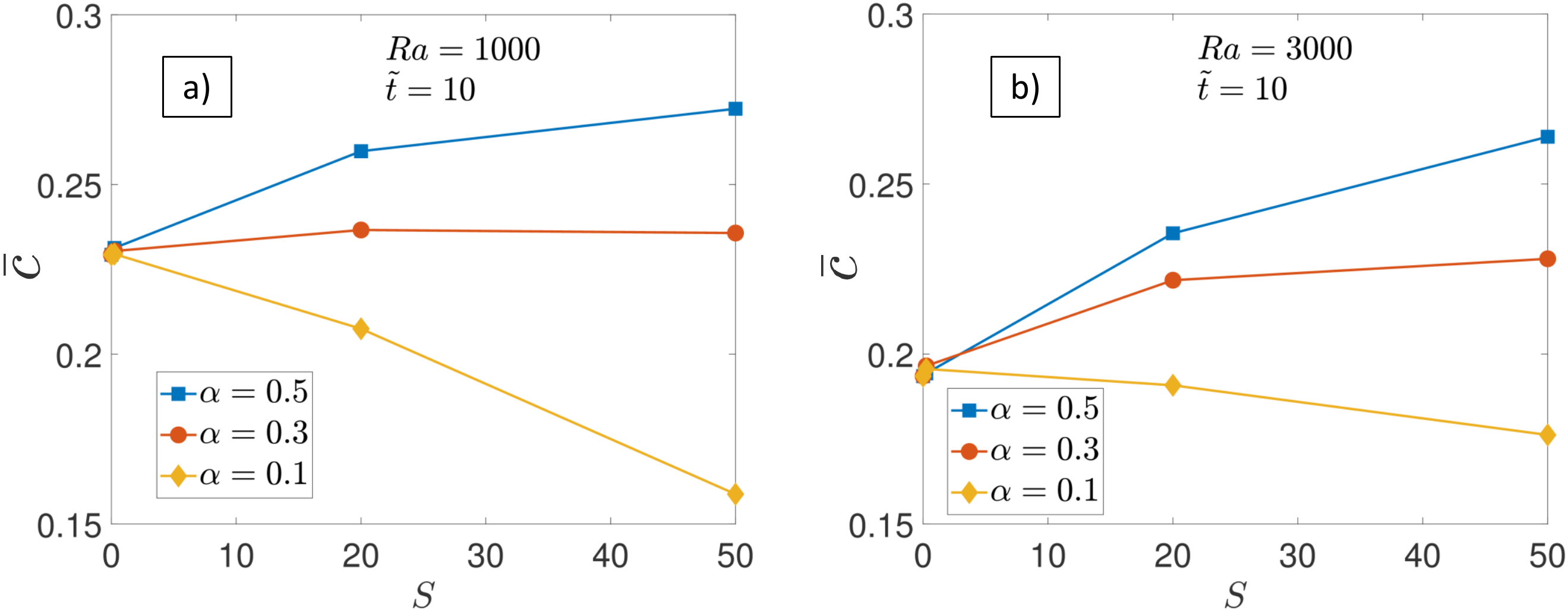}
\caption{\label{fig:dissol_limits} Dependence of the mean concentration on the dispersion strength $S$ at dimensionless time $\tilde{t}=10$ in the 2D geometry, for values of the dispersivity ratio $\alpha$ of $0.1$, $0.3$ and $0.5$, and  for a) $Ra=3000$ and $Ra=1000$.}
\end{figure}
\begin{figure}
\centering
\includegraphics[width=0.5\textwidth]{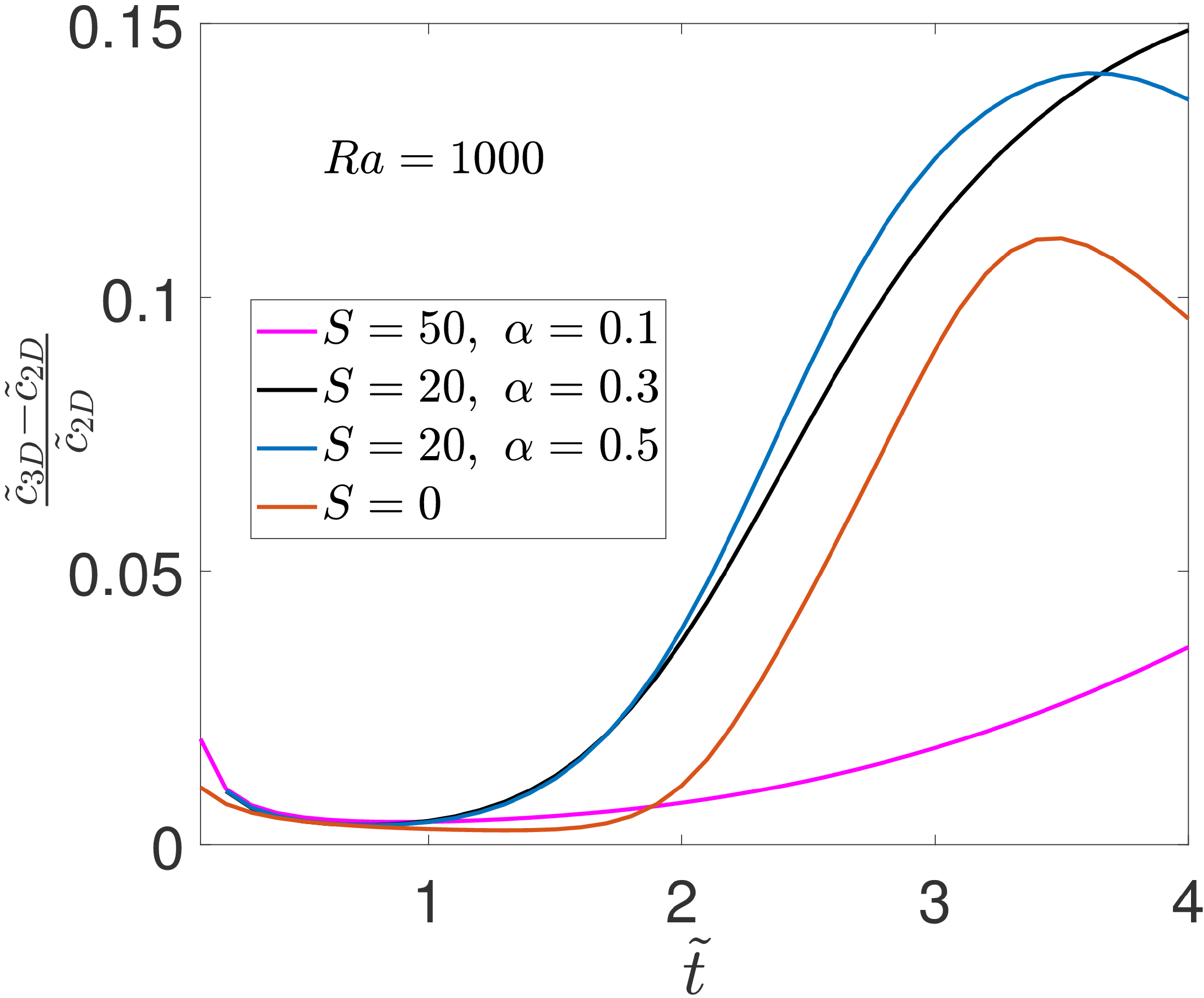}
\caption{\label{fig:Dissolution2Dvs3D} Relative difference between the temporal evolution of the concentration in the 2D geometry and in the 3D geometry, for different dispersion strengths and dispersivity ratios.}
\end{figure}
After the nonlinear regime is established, the evolution in the flux profile fluctuates widely, as previously discussed. Thus it is not obvious how the flux evolution with time translates into the evolution of the mean concentration $\bar{c}$ inside the flow domain, which is the most straightforward measure of the advancement of solubility trapping of CO$_2$ inside the geological formation.

Figure \ref{fig:Dissolution}a shows the evolution of $\bar{c}$ in the 2D geometry as a function of time for $Ra=1000$ and $Ra=3000$, $\alpha=0.1$ and $\alpha=0.5$, and values of $S$ ranging between $S=0$ and $S=50$. As seen previously on the onset time, at any given time an increase in $S$ for $\alpha=0.1$ leads to a decrease of the mean concentration, while for $\alpha=0.5$ it leads on the contrary to an acceleration of the dissolution. Increasing $\alpha$ also leads to a larger average growth rate. At the larger Rayleigh, the growth is also quicker than at the lower Rayleigh. These dependences on the Rayleigh and dispersion parameters are summarized in Fig.~\ref{fig:dissol_limits}, which shows $\bar{c}$ as a function of the dispersivity ratio for different dispersion strengths, both for $Ra=3000$ and $Ra=1000$. For $\alpha=0.1$ and $Ra=3000$ the dependence on $S$ is non-monotonic, and for $S\ge 20$ the dependence on $\alpha$ is also non-monotonic, for both $Ra=3000$ and 
$Ra=1000$. In summary, the mean concentration evolution seems to be consistent with, and, possibly, mostly controlled by, the behavior of the onset time: a configuration that exhibits a larger onset time will get a head start and keep in time.

Comparison of the data obtained in the 3D and 2D geometries (Fig.~\ref{fig:Dissolution2Dvs3D}) for $Ra=1000$ shows that the effective dissolution, as measured by the time evolution of $\bar{c}$, is somewhat faster in the 3D system (by less than 15\%), with a peak in the relative difference occurring at a time at which the 3D convection has already started while the 2D convection hasn't. Unfortunately our comparison addressed only non-dimensional times smaller than $4$, due to the high computational time of the 3D simulation.

\subsection{Scalar dissipation rates}

\begin{figure}
\centering
\includegraphics[width=1\textwidth]{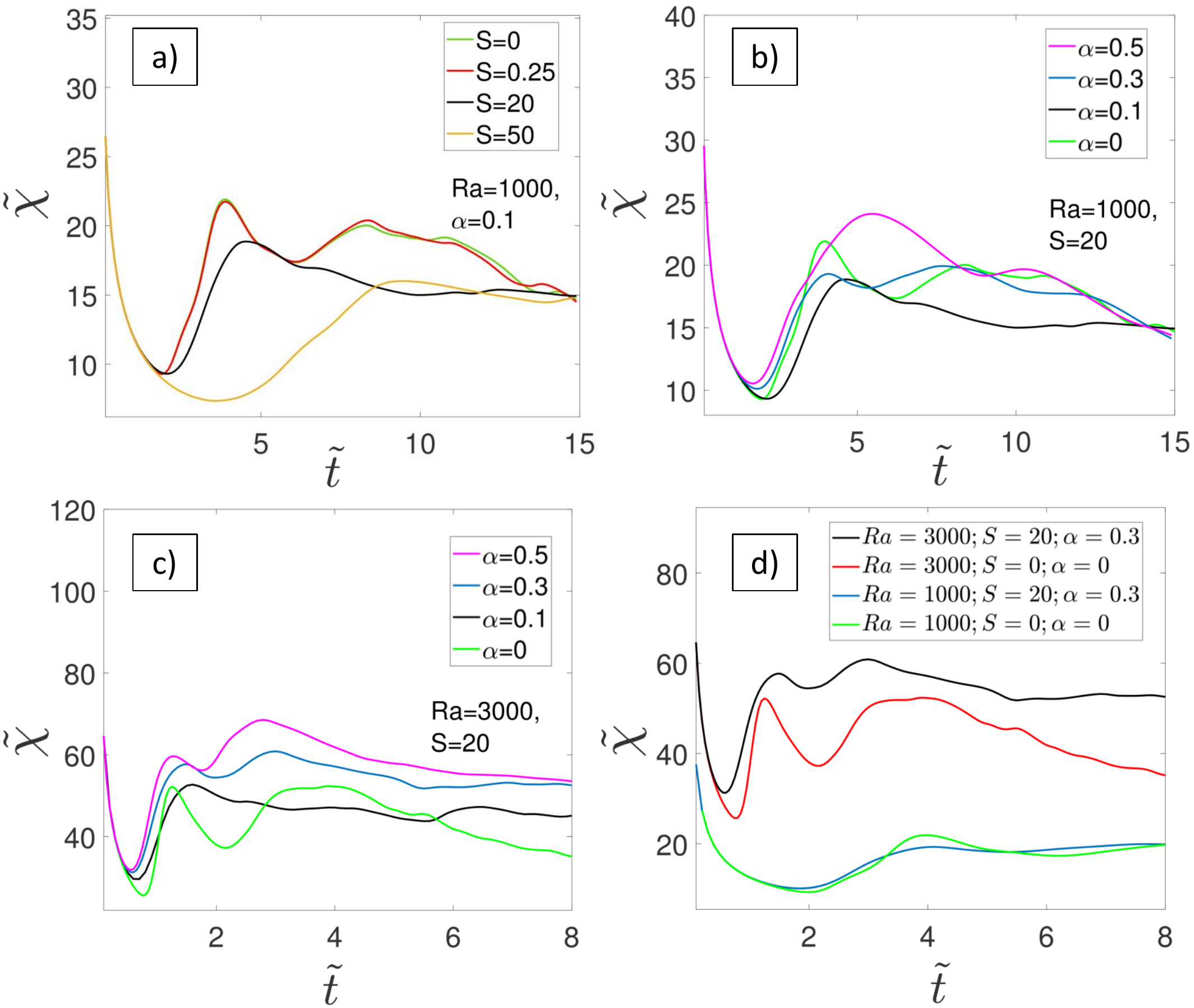}
\caption{\label{fig:SDR}Temporal evolution of the scalar dissipation rate in the 2D geometry a) for $Ra=1000$ and for different dispersion strengths $S$ with $\alpha=0.1$, and b) for $Ra=1000$ and for different dispersivity ratios $\alpha$ with $S=20$; c) is identical to b), exept for the value of the Rayleigh number, which is $Ra=3000$; d) shows a comparison between configurations which differ only by the Rayleigh number ($Ra=1000$ or $Ra=3000$).}
\end{figure}
\begin{figure}
\centering
\includegraphics[width=1\textwidth]{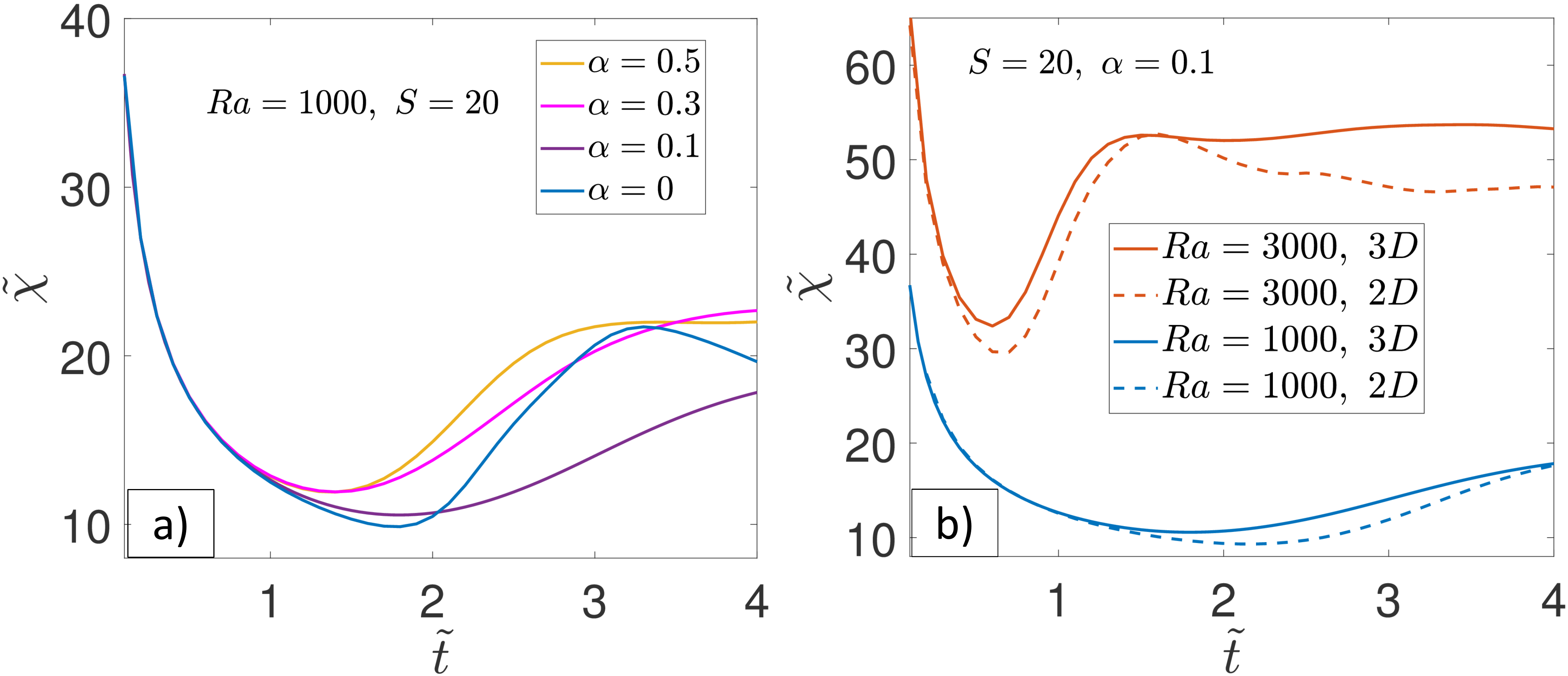}
\caption{\label{fig:SDR_3D} a) Same plot as that in Fig.~\ref{fig:SDR}b, but in the 3D geometry. 
b) Comparison between results obtained in the 2D and 3D geometries for $S=1$ and $\alpha=0.1$, with either $Ra=1000$ or $Ra=3000$.}
\end{figure}
The scalar dissipation rate (SDR) is a global measure of how strong  mixing is within the flow domain, which is controlled by the intensity of concentration gradients in the system. Fig.~\ref{fig:SDR} shows its temporal evolution in the 2D geometry, for different parameter configurations. In particular, Fig.~\ref{fig:SDR}a addresses the behavior of the scalar dissipation for different strengths $S$ of the dispersion,  with  $Ra=1000$ and $\alpha=0.1$; the configurations are exactly identical to those of Fig.~\ref{fig:Flux}a, and the different plots corresponding to the same configurations in the two figures exhibit a strong correlation.  In other words, the overall strength of mixing in the entire flow domain is strongly correlated with the dissolution flux at the top boundary of the domain, at least at this Rayleigh number, $Ra=1000$.  This means that, similarly to the dissolution flux, 
the SDR is observed to first decrease until the non-linear convection sets in, then increase sharply, and then fluctuates more or less strongly  around a stationary or slowly decreasing tendency. The time at which the mininum value of the SDR is observed, is directly related to the onset time defined from the minimum of the dissolution flux's, and its dependence on $Ra$, $S$ and $\alpha$ thus follows the same behavior as discussed above (section~\ref{sec:FluxOnset}). 
For a dispersivity ratio of $\alpha=0.1$ and at $Ra=1000$, introducing dispersion ($S>20$ vs. $S~0$) induces a weak decrease in the SDR (Fig.~\ref{fig:SDR}b), which is consistent with what was observed for the dissolution flux. For $S=20$ and at $Ra=1000$, an increase in the dispersivity ratio increases the SDR in the fluctuation regime as soon as $\alpha>0$ (Fig.~\ref{fig:SDR}b). However, a system with no dispersion seems to mix as well as the configuration with $S=20$ and $\alpha=0.3$. 
Fig.~\ref{fig:SDR}c shows the same dispersion configurations as Fig.~\ref{fig:SDR}b, except that the Rayleigh value is larger in the former ($Ra=3000$) than in the latter ($Ra=1000)$. Here, in contrast to what was observed for the dissolution rate, a larger Rayleigh induces a significantly larger SDR. This is confirmed by Fig.~\ref{fig:SDR}d, where the temporal evolution of the SDR is shown for the two investigated Rayleighs and for a dispersion that is either non-existent or defined by $S=20$ and $\alpha=0.3$.  We conclude that increasing the Rayleigh does increase the concentration gradients, as expected due to the larger density contrast between the CO$_2$-enriched solution and that which is devoid of CO$_2$, and thus strengthens the mixing, but also leads to stronger convection, which, by advection, tends to restrict the occurrence of strong concentration gradients to smaller regions at the boundaries between downward flow and upward flow; these two antagonistic effects seem to balance each other so that the dissolution rate in the fluctuating regime is not much influenced by the Rayleigh (see section~\ref{sec:FluxOnset}). But of course the Rayleigh number strongly impacts the onset time of convection, which we have discussed above (this is well known from the literature). 

Fig.~\ref{fig:SDR_3D}a is identical to Fig.~\ref{fig:SDR}a, except that it was obtained in the 3D geometry. It exhibits a behavior fully consistent with that of its 2D counterpart. In Fig.~\ref{fig:SDR_3D}b, the temporal evolution of the SDR are shown for $S=20$ and $\alpha=0.1$ and for the two investigated Rayleigh numbers, obtained either in the 2D or in the 3D geometry. These plots confirm that the dimensionality of space has little impact on the SDR, but that the SDR is all the larger as the Rayleigh is larger.

\subsection{Long term behavior of convective dissolution}

\begin{figure}
\centering
\includegraphics[width=0.8\textwidth]{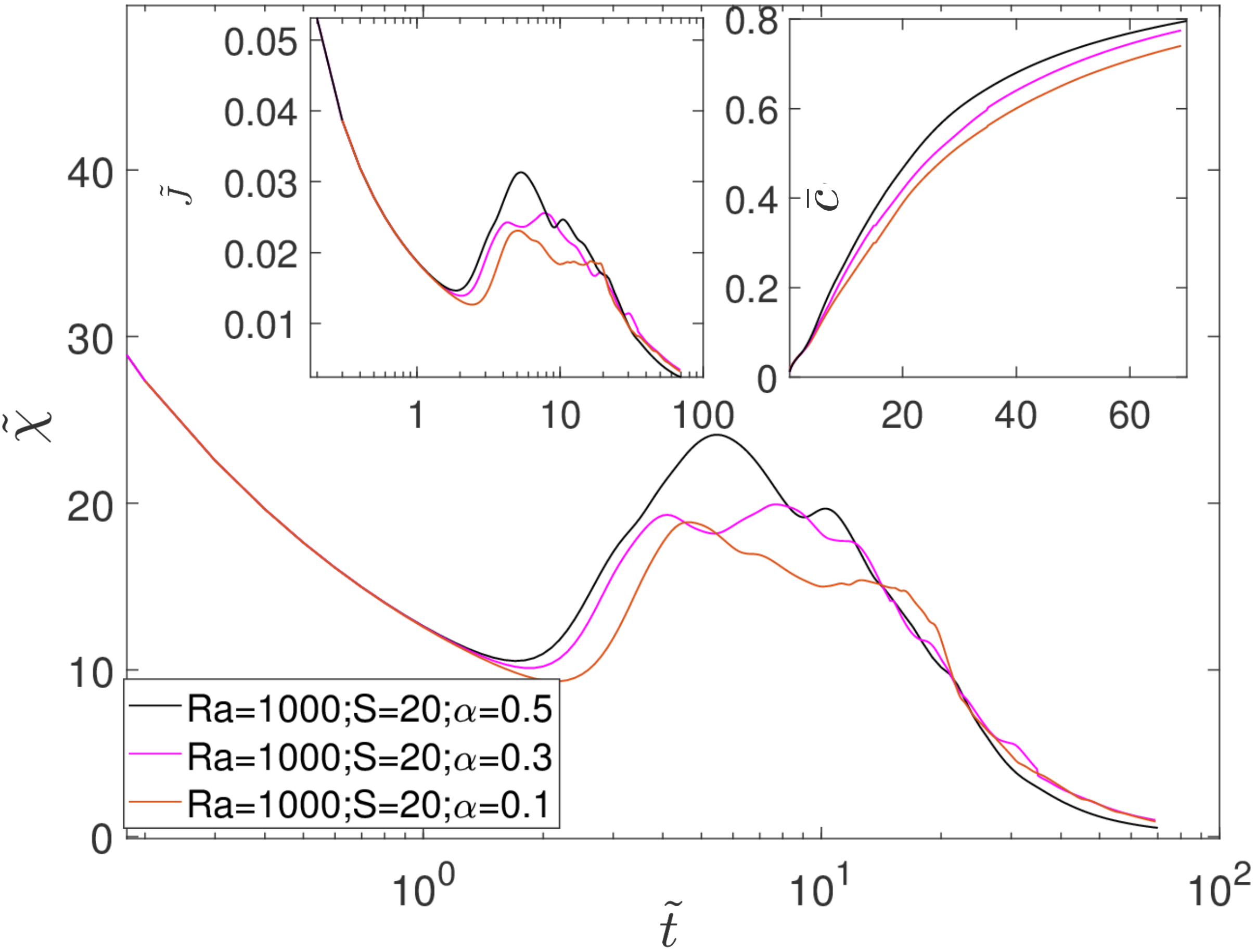}
\caption{\label{fig:long_times}Impact of the dispersivity ratio for $Ra=1000$ and $S=20$ on the scalar dissipation rate at large time, as well as on the total dissolution flux (left inset), and mean concentration (right inset). 
}
\end{figure}
We now examine how convective dissolution behaves in term of the scalar dissipation rate (Fig.~\ref{fig:long_times}), total dissolution flux (left inlet in Fig.~\ref{fig:long_times}), and mean concentration (right inlet in Fig.~\ref{fig:long_times}) at large times. These temporal evolutions over non-dimensional times as large as $70$ could only be obtained in the 2D geometry, due to obvious constraints of computational times. They are plotted for $Ra=1000$ and $S=20$ and 
for various dispersivity ratios. 
At large times (i.e., in the shutdown regime), the disparity due to differences in dispersivity ratios,  introduced in the dynamics after the initiation of convection, tend to diminish, as observed from the converging dissolution flux and scalar dissipation rates. This is expected since, as convection shut downs, fluid velocities decrease and thus render the impact of dispersion 
insignificant. Nevertheless, the impact of the dispersion on the onset time of convection and on the constant dissolution flux in the constant flux/fluctuation regime is clearly seen in the temporal evolution of the mean concentation, which is essentially a temporal primitive of the dissolution flux. The discrepancies in $\bar{c}$ are introduced in the constant flux regime and cease to increase at large times, but remain. 
The net dissolved mass depends montonicly on the dispersivity ratio and is largest for $\alpha=0.5$, which features the smallest non-linear onset time and the largest peak in the dissolution flux and scalar dissipation curves.  These conclusions remain valid across a range of different configurations of $Ra$ values and dispersion parameters, which are not shown here. 

\section{Discussion}
\label{sec:discussion}

\subsection{Impact of dispersion on convective dissolution}
\label{sec:roleOfDispersion}

\subsubsection{Summary  of our findings:}

From the results presented above, we can draw the following synthesis regarding the impact of hydrodynamic dispersion on convective dissolution:  
 
 1) The presence of mechanical dispersion, irrespective of the dispersivity ratio, works towards decreasing the finger number density (FND) of the ensuing fingers, at  any time and any vertical coordinate at which at least one finger exists. The penetration depth, as defined by an iso-concentration line (2D geometry) or surface (3D geometry) is also smaller with mechanical dispersion than with molecular diffusion alone. These two observations can both be related to the role of transverse dispersion, which favors transverse coalescence of concentration fingers, thus contributing to decreasing the FND, and  which tends to homogenize the concentration field, thus slowing down the convection which is driven by density contrasts resulting from concentration contrasts. Note however that for a given amplitude of the longitudinal dispersion, increasing the transverse dispersion has little impact on the finger number density. Such observations on the penetration depth have previously been made by \citet{menand2005dispersion}, while similar observations on the finger structure  have previously been reported experimentally by \citet{wang2016three}, who imaged 3D granular bead packs by X-ray tomography and used three different sizes of bead to vary the dispersivity length of the granular porous medium. 

Considering the same parameters as \citet{wang2016three} in order to compare our numerical results to their experimental measurements, i.e.  
$Ra=3000$, $K = 2.63\times10^{-10}$ m$^2$, $\phi = 0.5$, $g = 9.81$ m$^2\cdot$s$^{-1}$, $\eta = 0.001115$ Pa$\cdot$s, $\Delta \rho = 14$ kg$\cdot$m$^{-3}$, $D_0 = 2\times10^{-9}$ m$^2 \cdot$s$^{-1}$, $\alpha_\text{L} = \times10^{-3}$ m, $U_\text{ref} = K \Delta \rho g / \eta$, $S = \alpha_LU_\text{ref}/D=20$, $H = \eta \phi D Ra / (K \Delta \rho g)=11$ cm, we  obtain a FND of order $0.5$ N$\cdot$cm$^{-2}$ or less, in good agreement with the measurements presented in the experimental study. Although our simulations are performed with periodic boundary configuration, the effect of lateral confinement on the ensuing FND does not seem to strongly impact the result.

2)  Hydrodynamic dispersion slows down the establishment of the finger velocity as compared to molecular diffusion alone, again due to its transverse component which tends to decrease concentration contrasts/gradients. This translates in terms of the penetration length at a given non-dimensional time, which is less when hydrodynamic dispersion exists than without it. In the constant flux regime observed at sufficiently large Rayleigh, due to the large fluctuations around the long time tendency, it is not obvious from our current numerical data if the ratio of the transverse to the longitudinal dispersivity of the medium impacts the convection velocity.
 
3) Previous numerical studies on the impact of dispersion on convective dissolution \cite{hidalgo2009effect,ghesmat2011effect} have brought forth that the onset time decreases with the dispersion strength $S$. In contrast to these studies, we observe that the dependence of $t_\text{on}$ on $S$ is not necessarily monotonous, and that in addition it depends both on the Rayleigh number and on the dipersivity anisotropy $\alpha$. At dispersivity ratios typical of subsurface formations ($\alpha \simeq 0.1$), the onset time increases with  dispersion strength, which means that hydrodynamic dispersion slows down the development of convection. At larger dispersivity ratios, the onset times increases more slowly with $S$ and can even decrease with $S$ if $\alpha$ is sufficiently large (e. g., $\alpha=0.5$). Furthermore, for an intermediate value of $\alpha$ and for a sufficiently large Rayleigh, the dependence of the onset time on the dispersion strength can be non-monotonic, decreasing with $S$ for values of $S$ smaller than a crossover value, and increasing for values larger than the crossover value. In most natural systems as well as granular porous media prepared in the laboratory, however, the dispersivity ratio is close to $0.1$, so we should expect hydrodynamic dispersion to act to increase the onset time and thus act as a hindrance to convective dissolution. Note also that in the real world the dispersivity length scales with the typical grain size of the porous medium, $a$, and the Rayleigh number scales with the permeability, which scales as $a^2$. We discuss the link between our findings, those of previous numerical simulations, and those of experimental studies in section~\ref{sec:comparison_to_past_studies} below.

4) The dispersion strength also impacts the constant flux regime, which would be probably more well-established if the Rayleigh number were of order $10^4$ rather than 1000 as investigated here. Nevertheless, the dispersion strength has a clear effect on the mean flux, and that effect is strongly impacted by the dispersivity ratio: at $\alpha=0.1$ increasing the dispersion strength decreases the value slightly, while at $\alpha=0.5$ a much more significant increase of the mean flux value is observed when increasing $S$.

5) The combined impact of the onset time and mean plateau flux on the overall dispersion of CO$_2$ in the liquid phase is captured by the time evolution of the mean CO$_2$ concentration $\tilde{c}$. At a time well into the constant flux regime, the behavior of the mean concentration as a function of $S$ and $\alpha$ is fully consistent with that of the onset time, which shows that the main control on the overall dissolution is the onset time rather than the plateau value. This is consistent with our observations on the long time behavior convection dissolution, where the global dissolution is seen to be controlled by the time scale of the convection's initiation. The impact of the dispersivity ratio on the mean concentration, in particular, is exactly opposite to that on the onset time, as a larger onset time translates into a smaller mean concentration at any given time after the convection has kicked in. Thus, a higher $\alpha$ translates into a stronger convective dissolution. This can be attributed to the fact that stronger transverse dispersion corresponds to a greater capability of the fingers to mix with the resident fluid, as evidenced by the temporal evolution of the scalar dissipation rate. These findings are consistent with those of \citet{Marco2021current}, who also found that transverse dispersion is an important control factor in the early stages of dissolution. Considering the standard value of $0.1$ for the dispersivity ratio, however, leads to a decreasing behavior of the mean concentration as a function of the dispersion strength, at any given time  into the constant flux regime. Therefore, for most natural formations and porous media prepared in the laboratory, one would expect hydrodynamic dispersion to hinder the efficiency of convective dissolution. This is in stark contrast with the common view brought forth by previous numerical studies \citep{hidalgo2009effect,ghesmat2011effect}; we shall now discuss this discrepancy.

\subsubsection{How do our results compare to those of previous numerical and experimental studies ?}
\label{sec:comparison_to_past_studies}

The dependence of convective dissolution on the strength of dispersion $S$ has been a matter of debate in the literature, as discussed in the introduction. Generally, experimental findings (e.g., \citet{liang2018effect,menand2005dispersion}) have supported that dispersion acts towards strengthening convective dissolution,  whereas numerical investigations (e.g., \citet{ghesmat2011effect}) have shared an opposite view. As mentioned above, for many natural porous media the anisotropy ratio is close to $0.1$, for which our model predicts an increasing dependence of the onset time on dispersion strength. So why have previous numerical simulations predicted the contrary for similar configurations ? It is due to the way those numerical studies have non-dimensionalized the governing equations: it leads to their Rayleigh number being dependent on the dispersion strength, so that increasing the dispersion strength in the non-dimensionalized simulation (while maintaining the medium's porosity and permeability constant) results effectively in also increasing the Rayleigh number, which is  
$Ra^*=Ra/(1+S^*)$ with $S^*=S/(1+S)$, where $Ra$ and $S$ are the Rayleigh number and dispersion strength defined in the present study  (see more details in  Appendix~\ref{appA}). In that case the increases in $S$ and $Ra$ play antagonistic roles, but the increase in Rayleigh number, which tends to decrease the nonlinear onset time, easily dominates. On the contrary, in the present numerical study $S$ and $Ra$ are two independent parameters. More details on the impact of the non-dimensionalization scheme on the apparent role of dispersion are given in Appendix~\ref{appA}.

Few experimental studies have addressed the impact of hydrodynamic dispersion's intensity on convective dissolution. \citet{liang2018effect} have investigated convective dissolution with analog fluids in granular porous media consisting of glass beads, and measured a mean finger separation (proportional to the inverse of the FND) in the constant flux regime. By changing the size of the glass beads, they scaled the dispersivity lengths of the medium; they observed that the mean finger separation  increases as a function of the dispersion strength. Hence they found that the FND decreases with dispersion strength, which is exactly what we observe and is typically associated to a stronger convection resulting from a decrease in the onset time as a function of the hydrodynamic dispersion. This result is thus in contradiction to the numerical simulation of \citet{ghesmat2011effect}) and similar to our numerical findings. Can we then argue that our numerical simulation behave as an experimental study such as that of \citet{liang2018effect} is expected to ? They do qualitatively, by in fact they are not entirely comparable to such experiments, because increasing the typical grain size $a$ in an increase to increase $S$ does also increase the Rayleigh number $Ra$ because the transmissivity typically grows with $a^2$ (see e.g. the Kozeny-Carman model \citet{carman1939permeability}); to keep the Rayleigh number constant, the density contrast should be adjusted as well. So when simply changing the grain size, the dispersion strength cannot be increased without increasing the Rayleigh number at the same time, as is the case for the numerical simulation of \citet{ghesmat2011effect}. Our conclusion is that both approaches do not decouple the dispersion strength from the Rayleigh, in contrast to our present study, but that in the experimental study the effect of hydrodynamic increase dominates over that of the Rayleigh number increase, while in the numerical simulations the opposite occurs.

\subsection{Impact of space dimensionality}

 In the literature a vast majority of the studies of convective dissolution have been performed in 2D setups, either numerically or in experiments. In particular, numerical studies of the impact of dispersion have so far been performed in 2D, while the studies by \cite{wang2016three,liang2018effect} are to our knowledge the only ones that have addressed the topic experimentally. A handful of numerical studies have presented 3D simulations, but only one has compared 2D and 3D results:  \citet{Pau2010threeD} have shown that increasing the space dimensionality from 2D to 3D results in a slight decrease in onset time and a slight increase in mass flux; note however that these authors do not account for hydrodynamic dispersion in their model.  In fact, to our knowledge, only one 3D numerical investigation accounts for hydrodynamic dispersion in its model \cite{Erfani2021Reaction3D} , and it does not compare 2D and 3D results.

In this study we have systematically simulated convective dissolution for different sets of parameters ($Ra$, $S$, $\alpha$), both in the the two-dimensional (2D) and three-dimensional (3D) geometries, so as to compare the behaviors of the various observables between in 2D and 3D. Due to the heavy computational load associated with 3D simulations, the parameter space has not been describes as completely in 3D as in 2D, but in most cases we have been able to confront the 2D and 3D results. We find that in the absence of dispersion, the finger number density (FND), considered at $\tilde{t}=4$, 
is similar at the top of the flow domain between 3D and 3D geometries when $Ra=1000$, but is larger (by about $\sim 20$\%) in the 3D flow domain when $Ra=3000$. However, the FND reaches 0 at about the same vertical position as in the 2D flow domain, both for $Ra=1000$ and $Ra=3000$. In addition, the impact of dispersion on the vertical FND profiles is similar in 3D and 2D. Accordingly, the temporal evolution of the finger penetration, as defined by the $\tilde{c}=0.25$ isoline, is very similar in 3D and 2D, whether hydrodynamic dispersion is present or not and for the two investigated Rayleigh number values. The dependence of the onset time on the dispersion strength and dispersivity ratio is also similar in 2D and 3D, for the Rayleigh value for which it could be compared ($Ra=1000$), but the onset time is systematically smaller (by about $20$\%) in the 3D geometry, in agreement with the previous findings of \citet{Pau2010threeD}. The slight increase in plateau mass flux between 32D and 3D observed by these authors is also confirmed in our numerical simulations, both for $Ra=1000$ and for $Ra=3000$. 

In conclusion, it seems that the conclusions of \cite{Pau2010threeD} concerning the impact of space dimensionality, mentioned above, still hold independently of whether hydrodynamic dispersion impacts the convective dissolution process. There is however one qualitative difference between the 3D and 2D numerical simulations:  the value of the FND at the top domain boundary, which may differ between 2D and 3D, with a discrepancy depending on the value of the Rayleigh number.

\section{Conclusion}

We have developed a new three-dimensional continuum scale model of convective dissolution that accounts for anisotropic hydrodynamic dispersion. The OpenFOAM CFD package was used to implement the model in a numerical simulation which was then used it to investigate systematically the impact of dispersion strength ($S$) and anisotropy ratio ($\alpha$) on convective dissolution, both in two-dimensional (2D) and in three-dimensional (3D) geometries. We examined in detail how vertical profiles of the number density, the temporal evolution of the fingers' penetration depth, the nonlinear onset time, the dissolution flux after convection has developed, and the temporal evolution of the mean concentration and the scalar dissipation rate, are impacted by $S$ and $\alpha$; we also confronted the 2D and 3D findings in each case to see how much they differ. This was done for two Rayleigh numbers: $1000$ and $3000$.

We find that the overall impact of hydrodynamic dispersion on the efficiency of the convective dissolution process is mostly controlled by its impact on the onset time of convection. In contrast with previous numerical findings, which indicated a strong decrease of the onset time with dispersion strength for a standard value of the dispersivity ratio $\alpha=0.1$, we observe a clear but moderate increase of the onset time with $S$ for $\alpha=0.1$. This findings is consistent with those of the few experimental studies which have addressed the impact of dispersion, and tends to indicate that hydrodynamic dispersion is expected to be a hindrance to convective dissolution in subsurface formations. This discrepancy with previous numerical findings is due to their non-dimensionalization scheme, which led to the effective Rayleigh number (i.e., the relative strength of molecular diffusion with respect to buoyancy-triggered advection of the solute) being increased when the dispersion strength was increased; in our simulation the effective Rayleigh number is independent of the dispersion strength. We also observe that when increasing the dispersivity ratio the dependence of the onset time on the dispersion strength changes, with an increasing dependency for $\alpha=0.5$, and even a non-monotonic behavior at intermediate $\alpha$ values; this dependence $t_\text{onset}(S,\alpha)$ is actually also impacted by the Rayleigh number.

Comparison between the 3D and 2D results show that the 2D results are qualitatively similar to the 3D results on all accounts, except when it comes to the impact of the Rayleigh number on the finger number density at the top boundary of the domain. In particular, the results on the impact of the dispersion parameters $S$ and $\alpha$ on the dynamics of convective dissolution are qualitatively similar in 2D and 3D. 

This work thus provides a systematic assessment of the impact of hydrodynamic dispersion on convective dissolution in the context of the subsurface storage of CO$_2$. In doing so it also explains the discrepancy between the previous numerical and experimental studies of the impact of dispersion on the efficiency of convective dissolution. If opens prospects towards a precise mapping of the onset time's behavior as a function of the dispersion parameters and over a wide range of Rayleigh number values. Experimental studies in which the dispersivity lengths are varied independently of the Rayleigh number would be very useful to provide a closer comparison to our findings, as would experiments where the dispersivity ratio can be adjusted (it is possible by various means, see \citep{JacobTransverse2017,YangTransverse2021}). Note that recent pore scale experiments suggest that Darcy scale simulations may not be able to grasp the full complexity of the coupling between pore scale convection and solute transport \citep{brouzet2021SUBMITTED}; further developments may require  incorporating pore-scale effects under the purview of continuum-scale modelling in order to properly model the convection finger velocity in porous media without sacrificing the essential low-cost computational advantage provided by continuum simulations.

\section*{Acknowledgments}
\begin{acknowledgments}
This work was supported by ANR (Agence Nationale de la Recherche -- the French National Research Agency) under project {\em CO$_2$-3D} (grant nr. ANR-16-CE06-0001). We acknowledge enlightning discussions with J. Carrera Ram\'\i{}rez and J. J. Hidalgo during the course of the project. We also thank C. Soulaine and N. Mukherjee for suggestions regarding the numerical setup of the OpenFOAM model. 
\end{acknowledgments}

\appendix
\beginsupplement

\section{Validation of the numerical model}\label{appB}

\begin{figure}
\centering
\includegraphics[scale=0.25]{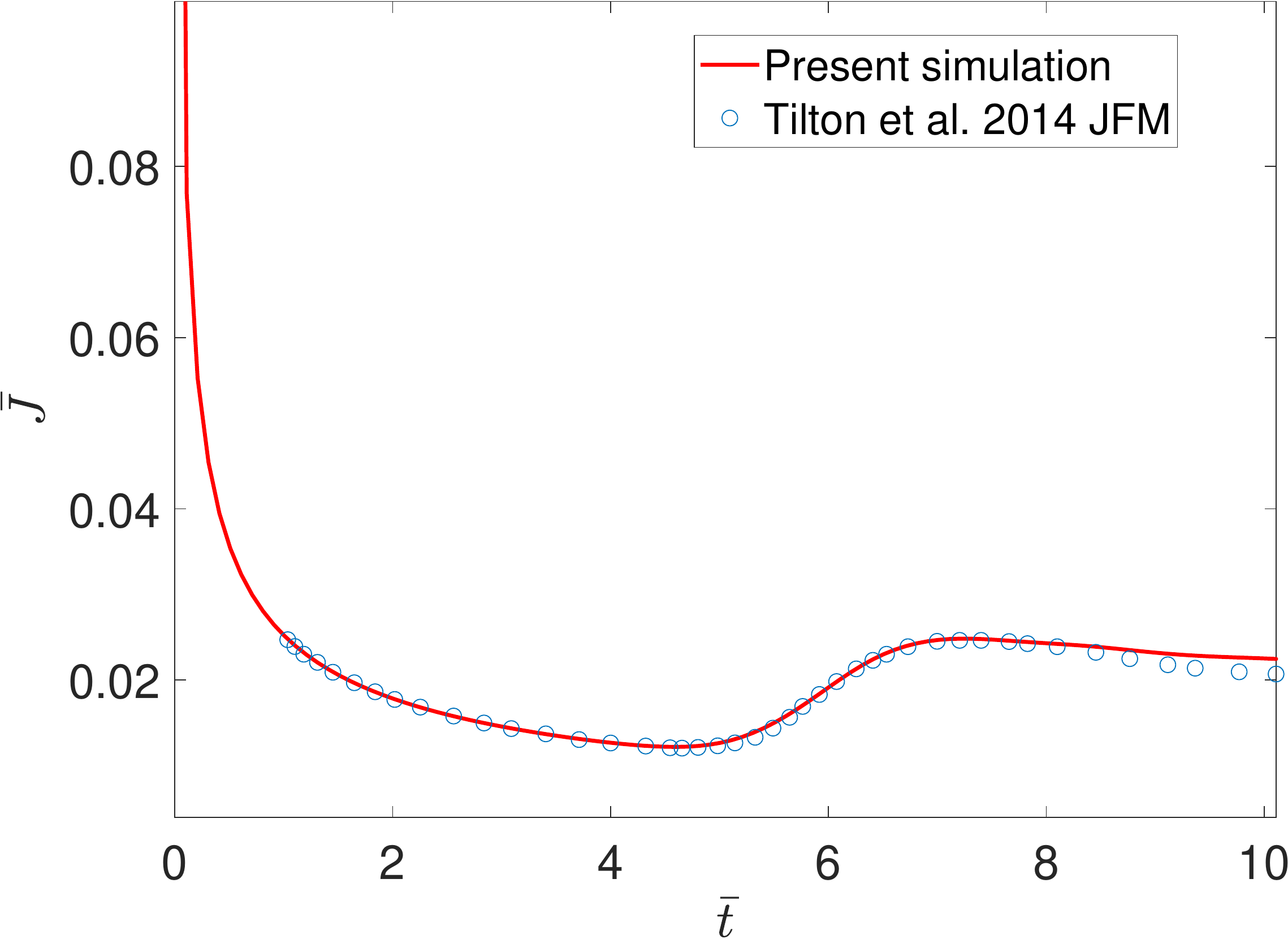}
\caption{\label{fig:verification} Temporal evolution of the solute flux a predicted by \citet{tilton2013initial}, and by our numerical simulation using the exact same parameters and no hydrodynamic dispersion.}
\end{figure}
To validate the numerical model, we have run our OpenFOAM simulation with the parameters considered by \citet{tilton2013initial}, and confronted our predicted temporal evolution to that predicted by the  numerical simulation of these authors, which used a spectral discretization method. As shown in Fig.~\ref{fig:verification} where the dimensionless flux, $\bar{J}$, is plotted as a function of the dimensionless time, $\bar{t}$, the agreement between the two predictions is excellent over the entire time range ($0<\tilde{t}\leq 10$). Note that $Ra=500$ and that  the dispersion parameters ($S$ and $\alpha$) was set to 0 in our simulation, as \citet{tilton2013initial} did not consider hydrodynamic dispersion.

\section{Spatial distributions of horizontal and vertical components of the velocity}
\label{sec:AppVxVz}

The maps of the components of velocity $\tilde{u}_x$ and $\tilde{u}_y$ in the two-dimensional geometry addressed in Fig.~\ref{fig:conc_vel_profile}, show that the horizontal velocities are largest at the top boundary of the domain, where the vertical velocity component is necessarily zero, and emphasize the finger structure in the $\tilde{u}_y$ maps. 
\begin{figure}
\centering
\includegraphics[width=\textwidth]{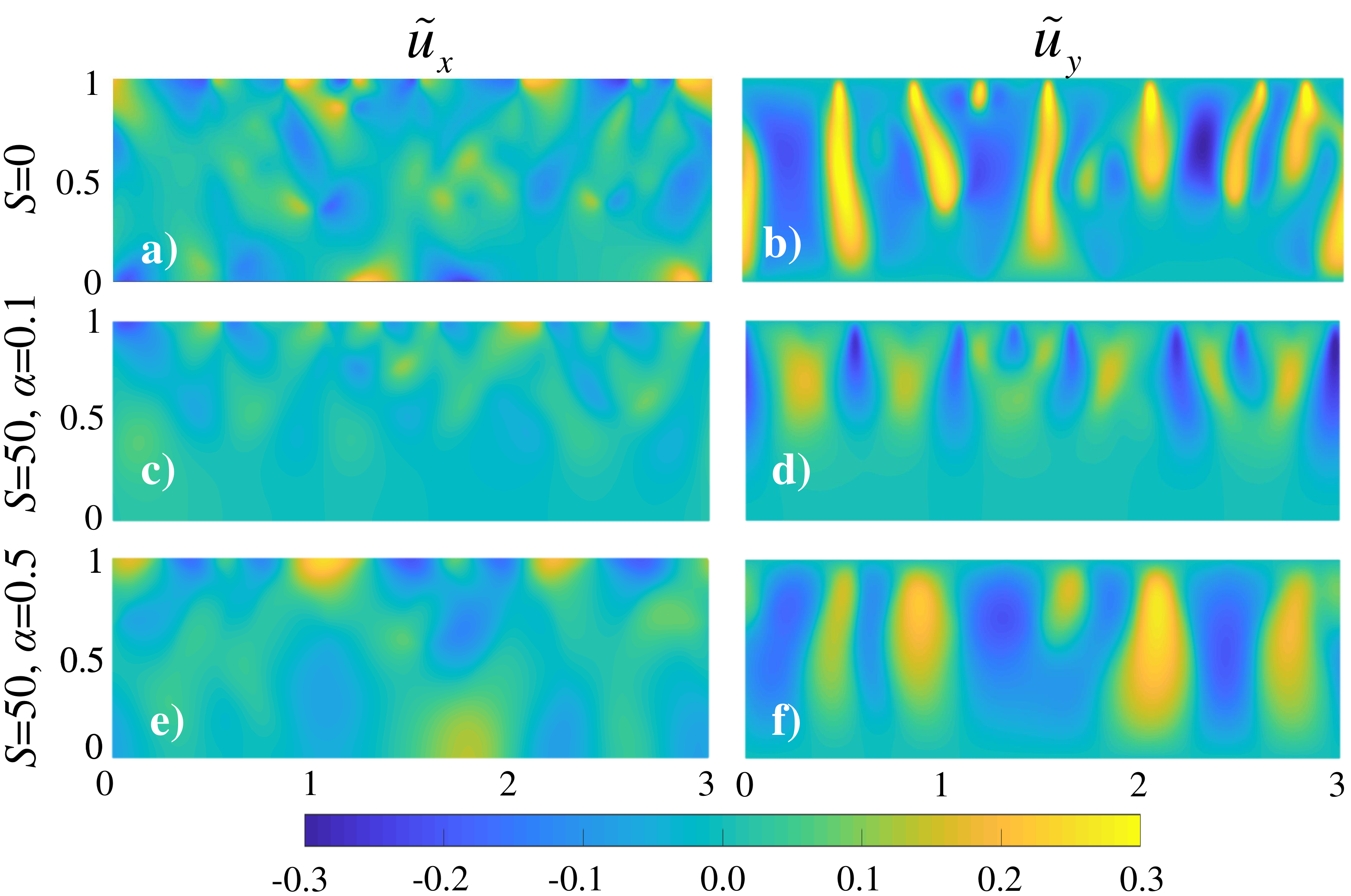}
\caption{\label{fig:Vx_and_Vy} Horizontal and vertical components of the 2D velocity field corresponding to the left column of the bottom panel of Fig.~\ref{fig:conc_vel_profile}, i.e., for $Ra=1000$ and for three dispersion configurations indicated on the left of the velocity maps: $S=0$, $S=50$ with $\alpha=0.1$, and $S=50$ with $\alpha=0.5$.}
\end{figure}

\section{On non-dimensionalization strategies}\label{appA}

\begin{table}
\centering
\caption{\label{tab:parameters2} Values of the physical parameters for the simulation illustrated in Fig.~\ref{fig:fndAnomaly}.}
\smallskip
\begin{tabular}{l | c}
{\bf Physical parameter} & {\bf Parameter value} \\
\hline \hline
Medium permeability ( $\kappa$) &  $10^{-10}$~m$^2$ \\
Medium porosity ($\phi$) &  0.3 \\
Viscosity ($\eta$) &  $0.001$ Pa$\cdot$s\\
Diffusion coefficient ($D_0$) & $10^{-9}$~m$^2$/s\\
Density difference ($\Delta \rho$) &  $10$~kg/m$3$\\
Longitudinal dispersivity length ($\alpha_\text{L}$) & $2.25 \cdot 10^{-2}$ m ($S=20$) and $2.8 \cdot 10^{-4}$ m (for $S=0.25$)\\
Ratio of dispersivity lengths ($\alpha$) & $0.1$\\
Channel height ($H$) & $1$~m\\
Channel length ($L$)  &  $3$~m\\
Initial concentration at top ($c_0$) &  1 arb. unit
\end{tabular}
\end{table}

\begin{figure}
\centering
\includegraphics[width=0.99\textwidth]{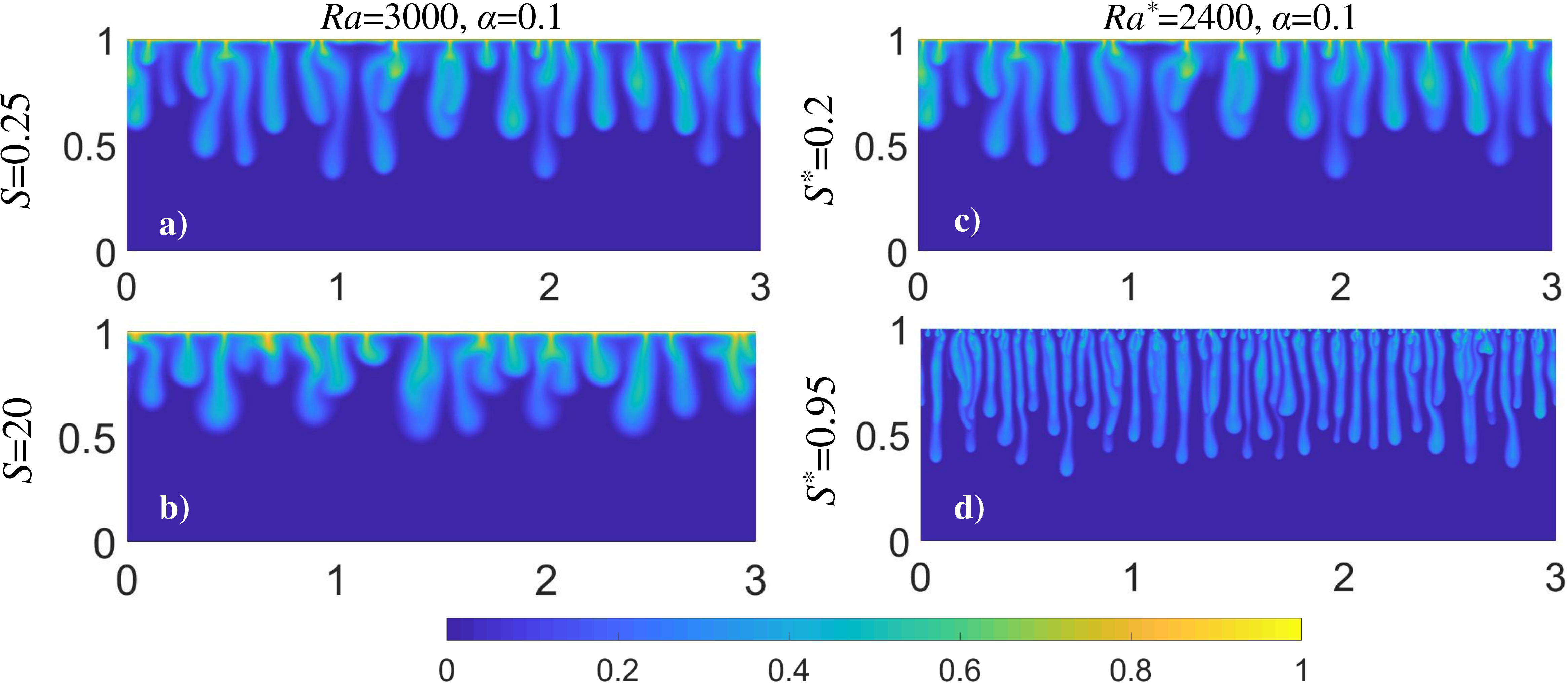}
\caption{\label{fig:fndAnomaly}(a, b): Snapshots of the concentration fields obtained at non-dimensional time $\bar{t}=5$ from the governing equations non-dimensionalized using our scheme (Eq.~\eqref{eq:nondim_eqs} and \eqref{eq:nondim_disp_accurate}), for a Rayleigh number $Ra=3000$ and a dispersion tensor defined by  $\alpha=0.1$ and (a) $S=0.25$ or (v) $S=20$; the dimensional parameters are given in Table~\ref{tab:parameters2}. (c): Same as (a), but from governing equations non-dimensionalized using the scheme of \citet{ghesmat2011effect} (Eq.~\eqref{eq:nondim_eqs} and \eqref{eq:nondim_disp_accurate_App}) for at Rayleigh number $Ra *=2400$ and for a dispersion tensor defined by $\alpha=0.1$ and $S^*=0.8$. (d) Same as (c), but for $S^*=0.95$.}
\end{figure}

Our findings on the role of hydrodynamic dispersion in convective dissolution are consistent with a number of previous experimental observations \citep{wang2016three,liang2018effect,menand2005dispersion}. The apparent contradiction between these findings and a number of previous numerical predictions \citep{hidalgo2009effect,ghesmat2011effect} can be explained by the non-dimensionalization scheme used in theses studies. They consider the same non-dimensional governing equations as used in the present study, namely Eq.~\eqref{eq:nondim_eqs} (in the case of \citep{hidalgo2009effect} they are slightly different as rock matrix compressibility is taken into account), but they non-dimensionalize the dispersion tensor in the following manner:
\begin{equation}
\label{eq:nondim_disp_accurate_App}
\overline{\mathbf{D}}=(1-S^*+S^*\alpha \| \mathbf{u} \|) \mathbf{I}+S^*(1-\alpha) \frac{\overline{u_{i}} \overline{u_{j}}}{\|\mathbf{u}\|} ~,
\end{equation}
where $\alpha$ is the dispersivity ratio as defined in the present study, and $S^*=(\alpha_\text{L} u_{\text{ref}})(D_0 + \alpha_L u_{\text{ref}})$ is the dispersion strength, which takes values in the range $[0,1]$. This equation can be compared to Eq.~\eqref{eq:nondim_disp_accurate} in our non-dimensionalization scheme, where the $1-S^*$ term in the factor of $\mathbf{I}$ in Eq.~\ref{eq:nondim_disp_accurate_App} above is replaced by $1$, and the dispersion strength $S=(\alpha_\text{L} u_{\text{ref}})/D_0$ takes values in the range $[0;+\infty]$. Obviously the two parameters accounting for the dispersion's strength are related to each other through $S=S^*/(1-S^*)$  or $ S^*=S/(1+S)$. Consequently, when using Eq.~\eqref{eq:nondim_disp_accurate_App} together with Eq.~\eqref{eq:nondim_eqs} to describe the coupled flow and solute transport dynamics,  the Rayleigh number $Ra^*$ which \citet{ghesmat2011effect} define must be $Ra (1-S^*)=Ra/(1+S)$, since the dimensional equations which they considered are identical to Eq.~\eqref{eqn:one}. In other words, $Ra^*$ is defined using an effective diffusion coefficient $D_0 + \alpha_\text{L} u_\text{ref}$. Hence the dispersion strength $S^*$ (or, equivalently, $S$) cannot be varied independently of the Rayleigh number $Ra^*$ : with any increase in dispersion strength, $Ra^*$ decreases. On the contrary, the Rayleigh number $Ra$ (the one we use) is independent of $S^*$ or $S$.

Note that \citet{wen2018rayleigh} have mentioned this issue with the non-dimensionalization scheme in \citep{ghesmat2011effect}. To illustrate its effect we show in Figure~\ref{fig:fndAnomaly} how increasing the dispersion strength impacts the convective fingers in a simulation that is based on our non-dimensionalized equations \eqref{eq:nondim_eqs} and \eqref{eq:nondim_disp_accurate} (Fig.~\ref{fig:fndAnomaly}a and \ref{fig:fndAnomaly}b), and in a simulation based on the non-dimensionalization scheme defined by Eq.~\eqref{eq:nondim_eqs}  and \eqref{eq:nondim_disp_accurate_App} (Fig.~\ref{fig:fndAnomaly}c and \ref{fig:fndAnomaly}d). Fig.~\ref{fig:fndAnomaly}a shows the concentration field obtained at non-dimensional time $\tilde{t}=5$ for a dimensional configuration defined by the dimensional parameters of Table~\ref{tab:parameters2}; the Rayleigh number is $Ra=3000$, the dispersion strength $S=0.25$, and the dispersivity ratio $\alpha=0.1$. Fig.~\ref{fig:fndAnomaly}b shows the same concentration field for the same configuration as in Fig.~\ref{fig:fndAnomaly}b, except that the dispersion strength is now $S=20$. As discussed at length in section~\ref{sec:results}, the solute fingers are thicker and penetrate more slowly into the flow domain when $S=20$ than when $S=0.2$ (compare  Fig.~\ref{fig:fndAnomaly}b to \ref{fig:fndAnomaly}a). 
Fig.~\ref{fig:Dissolution2Dvs3D} was obtained with the same domain size, medium properties and fluid properties as in Fig.~\ref{fig:fndAnomaly}a,  but using the second non-dimensionalization scheme with $Ra^*=2400$, $S^*=0.2$, and $\alpha=0.1$; in terms of our non-dimensional parameters, this corresponds to $Ra=3000$, $S=0.25$, and $\alpha=0.1$, i.e., to the exact same dimensional system as that of Fig.~\ref{fig:fndAnomaly}a; hence the concentration field is exactly identical to that in the latter subfigure. Fig.~\ref{fig:fndAnomaly}d was obtained with the same parameters as  Fig.~\ref{fig:fndAnomaly}c, except that the dispersion strength was increased to $S^*=0.95$ (corresponding to $S=20$ as in Fig.~\ref{fig:fndAnomaly}b). We see that with this non-dimensionalization scheme the finger have become much more narrow and have penetrated farther into the flow domain when increasing the dispersion strength; the reason is that, while $Ra^*$ has been kept constant between Fig.~\ref{fig:fndAnomaly}c and Fig.~\ref{fig:fndAnomaly}d), the effective Rayleigh number $Ra$ has changed and is equal to $48000$ in Fig.~\ref{fig:fndAnomaly}d, a much larger value for which the convection is much stronger. Conversely, for Fig.~\ref{fig:fndAnomaly}d to display the same concentration field as Fig.~\ref{fig:fndAnomaly}b, $Ra*$ would have to be set to $142.86$.

This explains why \citet{hidalgo2009effect} and \citet{ghesmat2011effect} have observed a very strong decrease of the nonlinear onset time as a function of the dispersion strength: increasing $S^*$ strongly increases the effective Rayleigh number $Ra$, which speeds up the onset of convection very much.


\end{document}